\documentclass[preprint,11pt]{elsarticle}
\usepackage[top=0.75in, bottom=0.75in, left=0.75in, right=0.75in]{geometry}

\usepackage{lineno}
\usepackage{graphicx}
\usepackage{amssymb}
\usepackage{amsmath}
\usepackage{multirow}
\usepackage{caption} 
\usepackage{booktabs}
\usepackage{longtable}
\usepackage{array}
\usepackage{color}

\journal{Mathematical Biosciences}

\begin{document}

\begin{frontmatter}

\title{Combination therapy for colorectal cancer with anti-PD-L1 and cancer vaccine: A multiscale mathematical model of tumor-immune interactions}

\author[label1]{Chenghang Li} 

\author[label2]{Haifeng Zhang} 

\author[label3]{Xiulan Lai\corref{cor}} 

\author[label1,label4]{Jinzhi Lei\corref{cor}} 

\affiliation[label1]{organization={School of Mathematical Sciences},
            addressline={Tiangong University}, 
            city={Tianjin},
            postcode={300387}, 
            country={China}}
        
\affiliation[label2]{organization={School of Mathematical Sciences},
	addressline={Jiangsu University}, 
	city={Zhenjiang},
	postcode={212013}, 
	country={China}}

\affiliation[label3]{organization={School of Mathematics},
	addressline={Renmin University of China}, 
	city={Beijing},
	postcode={100872}, 
	country={China}}

\affiliation[label4]{organization={Center for Applied Mathematics},
	addressline={Tiangong University}, 
	city={Tianjin},
	postcode={300387}, 
	country={China}}

\cortext[cor]{Corresponding author. E-mail address: xiulanlai@ruc.edu.cn (X. Lai), jzlei@tiangong.edu.cn (J. Lei)}

\begin{abstract}
The tumor-immune system plays a critical role in colorectal cancer progression. Recent preclinical and clinical studies showed that combination therapy with anti-PD-L1 and cancer vaccines improved treatment response. In this study, we developed a multiscale mathematical model of interactions among tumors, immune cells, and cytokines to investigate tumor evolutionary dynamics under different therapeutic strategies. Additionally, we established a computational framework based on approximate Bayesian computation to generate virtual tumor samples and capture inter-individual heterogeneity in treatment response. The results demonstrated that a multiple low-dose regimen significantly reduced advanced tumor burden compared to baseline treatment in anti-PD-L1 therapy. In contrast, the maximum dose therapy yielded superior tumor growth control in cancer vaccine therapy. Furthermore, cytotoxic T cells were identified as a consistent predictive biomarker both before and after treatment initiation. Notably, the cytotoxic T cells-to-regulatory T cells ratio specifically served as a robust pre-treatment predictive biomarker, offering potential clinical utility for patient stratification and therapy personalization.
\end{abstract}




\end{frontmatter}

\section{Introduction}	

Colorectal cancer (CRC) is a prevalent malignancy of the digestive system \cite{Siegel.CACancerJClin.2023}. In CRC treatment, surgical resection combined with adjuvant chemoradiotherapy has long been established as the standard therapeutic paradigm \cite{Siegel.CACancerJClin.2023,Biller.JAMA.2021}. Recently, advances in tumor immunology have reshaped the colorectal cancer therapeutic landscape \cite{Andre.NEnglJMed.2024,Diaz.LancetOncol.2022}. The combination therapy of immune checkpoint blockade with emerging cancer vaccines is improving CRC treatment outcomes \cite{Liu.NatCancer.2022}. This innovative therapeutic strategy synergistically activates the immune response of tumor-specific T cells. In this study, we developed a multiscale mathematical model to mechanistically elucidate the synergistic mechanisms of combined immune checkpoint blockade and cancer vaccine therapy in colorectal cancer, and to systematically evaluate anti-tumor efficacy, immune cell infiltration dynamics, and potential predictive biomarkers under different treatment strategies.

Immune checkpoint blockade therapy reactivates anti-tumor immune responses by targeting inhibitory signaling pathways such as the PD-1/PD-L1 axis, which triggers immunosuppression and leads to T cell exhaustion \cite{Wei.CancerDiscov.2018, Okazaki.NatImmunol.2013, Morad.Cell.2022}. Anti-PD-L1 antibodies specifically bind to PD-L1 molecules, thereby reversing tumor-mediated T cell dysfunction and restoring their normal activation state and effector functions \cite{Yamaguchi.NatRevClinOncol.2022}. On the other hand, cancer vaccines induce precise anti-tumor immune responses through the combined delivery of tumor-specific antigens and immune adjuvants \cite{Finn.NatRevImmunol.2003, Lin.NatCancer.2022}. The core mechanism involves two synergistic processes: (1) dendritic cells internalize, process, and present tumor-specific antigens via MHC molecules to activate antigen-specific T cells, and (2) immune adjuvants enhance T cell activation by boosting antigen immunogenicity and promoting dendritic cell maturation \cite{Finn.NatRevImmunol.2003,Lin.NatCancer.2022,Overwijk.CurrOpinImmunol.2017}. Together, they elicit robust anti-tumor immunity and establish long-term immunological memory.

Mathematical modeling has emerged as a pivotal tool for deciphering tumor-immune interactions \cite{Eftimie.BullMathBiol.2016,Eftimie.BullMathBiol.2023,Li.CSIAM-LS.2025}. Chen et al. \cite{Chen.MathBiosci.2022} developed an ordinary differential equation (ODE) model to reveal that the gut microbiome influences the anti-tumor efficacy of immune checkpoint inhibitors by regulating the host immune response. Liao et al. \cite{Liao.MathBiosci.2024} established a mechanistic model to explore the synergistic effects of combination therapy with radiation and anti-PD-L1 for tumor treatment. Lai et al. \cite{Lai.PNAS.2018} constructed a partial differential equation (PDE) model of cell-cytokine interactions, quantifying the combined efficacy of BET inhibitors and immune checkpoint inhibitors. Friedman et al. \cite{Friedman.BullMathBiol.2020} designed a PDE model to investigate the dynamics of tumor drug resistance and recurrence under BRAF inhibitor therapy. Zhang et al. \cite{ZhangRM.IJB.2025} established a novel hybrid multiscale model, systematically analyzing the regulatory mechanisms of multiple factors such as chemotherapy, PI3K inhibitors, and psychological stress on glioma growth.

In recent years, quantitative studies of CRC have grown progressively. Paterson et al. \cite{Paterson.PNAS.2020} developed a stochastic mathematical model of CRC initiation, validating the hypothesis that mutational selection within the APC-TP53-KRAS pathway dominates tumor evolutionary trajectories. Haupt et al. \cite{Haupt.PLoSComputBiol.2021} constructed a dynamical system with Kronecker structure, and quantitatively analyzed the multi-pathway co-evolutionary mechanism of CRC development. Furthermore, Mohammad-Mirzaei et al. \cite{Mohammad-Mirzaei.iScience.2023} demonstrated the pivotal role of macrophages in the CRC microenvironment using the PDE model, showing that macrophage polarization remodels immune cell distribution patterns and modulates T cell activity. Recently, Li et al. \cite{Li.NPJSystBiolAppl.2025} developed a quantitative cancer immune cycle model to systematically resolve the cross-scale tumor-immune system interaction network through a multi-compartmental system modeling approach. This study captures the inter-individual differences in the treatment of advanced CRC patients. 

Virtual sample generation technology integrates multi-source data and quantitative models to construct dynamically evolving virtual cohorts tailored to individual profiles \cite{Li.NPJSystBiolAppl.2025,Li.BullMathBiol.2024,Wang.JImmunotherCancer.2021,Anbari.NPJSystBiolAppl.2024}. In this framework, mathematical models serve as a reliable and interpretable foundation for mechanistic representation. Nevertheless, conventional mathematical approaches have predominantly characterized the average dynamics of tumor evolution, largely overlooking inter-individual heterogeneity. Recent efforts have incorporated stochastic parameter sampling to represent tumor variability. These methods often fail to systematically recover biologically plausible parameter configurations. This limitation substantially compromises the physiological credibility of such models. To overcome these challenges, we developed an approximate Bayesian computation (ABC) inference framework enhanced with a Gaussian kernel weighting strategy. This approach enables the efficient identification of parameter sets that simultaneously achieve high fidelity to empirical data and biological plausibility across high-dimensional parameter spaces. Thereby, it establishes a rigorous mathematical foundation for generating mechanistically interpretable virtual cohorts.

In this study, we developed a multiscale mathematical model to analyze the complex dynamics of tumor-immune interactions under anti-PD-L1 therapy and cancer vaccine treatment. Computational results revealed that multiple low-dose anti-PD-L1 administrations significantly reduced advanced tumor burden compared to baseline treatment, whereas maximal-dose cancer vaccine regimens demonstrated superior efficacy in controlling tumor progression. Furthermore, we employed approximate Bayesian computation (ABC) to generate virtual patient cohorts, effectively capturing interpatient heterogeneity in treatment responses. Our findings indicate that a modeling framework that integrates immune heterogeneity enables more accurate characterization of tumor evolutionary dynamics. Additionally, we analyzed immune cell distribution patterns under different treatment strategies, revealing that combination therapy significantly enhanced tumor-infiltrating cytotoxic T cell levels while suppressing regulatory T cell populations. We also discovered that cytotoxic T cells were a significant predictive biomarker both pre- and post-treatment, and the cytotoxic T cells-to-regulatory T cells ratio was an important predictive biomarker prior to treatment.

\section{Mathematical model}\label{sec2}

\begin{figure}[h!]
	\centering
	\includegraphics[width=14cm]{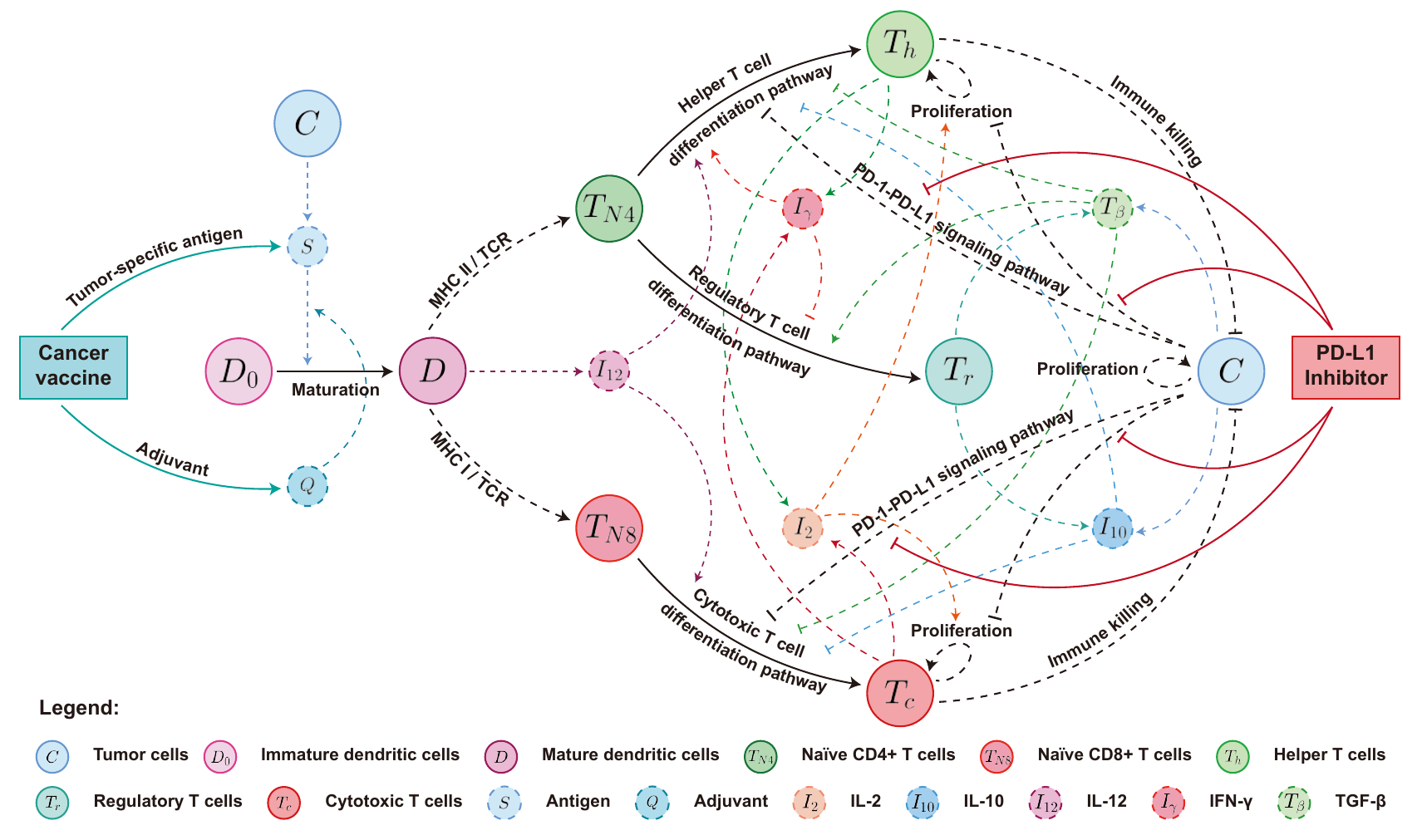}
	\caption{\textbf{Framework flowchart of the dynamic regulatory network.} Immune response initiation depends on tumor-specific antigens released by tumors and provided by cancer vaccines. Vaccine adjuvants enhance dendritic cell maturation and significantly improve their antigen-presenting capacity. Na\"{i}ve T cells recognize tumor-specific antigens presented by dendritic cells through TCR-pMHC binding. Driven by various cytokines, na\"{i}ve T cells differentiate into effector T cell subsets, which subsequently mediate tumor cell apoptosis through direct contact or cytokine secretion. Tumor cells exert immunosuppressive effects via PD-L1/PD-1 interactions with T cells. Solid black lines represent changes in cell state. Dashed black lines represent intercellular interactions. Colored dotted lines represent cytokine production and the action mechanism. Indigo and red colors represent the action mechanisms of cancer vaccines and PD-L1 inhibitors, respectively. Arrows denote promotion, proliferation, or activation, while blocking arrows indicate killing, blocking, or inhibition.}
	\label{Fig1}
\end{figure}

In the previous study \cite{Li.BullMathBiol.2024}, we developed a mathematical model to investigate the combination efficacy of immune checkpoint inhibitors and targeted inhibitors. On this basis \cite{Li.BullMathBiol.2024}, we further established a more critical dynamic regulatory network of tumor-immune interactions to investigate the effects of cancer vaccines and immune checkpoint inhibitors on the tumor evolution dynamics (Fig. \ref{Fig1}). The network operates across two distinct timescales: (1) fast-timescale dynamics governing cytokine production/degradation, and (2) slow-timescale dynamics regulating cell-cell interactions. This temporal separation highlights fundamental mechanistic differences between molecular and cellular processes. The units are as follows: cells are expressed in cells, cytokines in ng mL$^{-1}$, surface proteins in nmol L$^{-1}$, and antigens and adjuvants in $\mu$g L$^{-1}$. Model variables are summarized in Table \ref{Tab:Variable}.

\begin{table}[h!]
	\tiny
	\centering
	\caption{List of variables.}
	\label{Tab:Variable}
	\renewcommand\arraystretch{0.75}
	\begin{tabular}{>{\centering\arraybackslash}p{1cm}|>{\centering\arraybackslash}p{6cm}|>{\centering\arraybackslash}p{1.5cm}}
		\toprule
		{\bf Variate} & {\bf Description} & {\bf Unit} \\
        \midrule
		$D_0$    & The number of immature dendritic cells  &  $\mathrm{cells}$  \\
		$D$    & The number of mature dendritic cells   &  $\mathrm{cells}$  \\
		$T_{N4}$    & The number of na\"{i}ve CD4+ T cells   &  $\mathrm{cells}$  \\
		$T_h$  & The number of helper T cells   &  $\mathrm{cells}$  \\
		$T_r$  & The number of regulatory T cells   &  $\mathrm{cells}$  \\
		$T_{N8}$    & The number of na\"{i}ve CD8+ T cells   &  $\mathrm{cells}$  \\
		$T_c$  & The number of cytotoxic T lymphocytes   &  $\mathrm{cells}$  \\
		$C$    & The number of tumor cells      &  $\mathrm{cells}$  \\ 
		\midrule
		$I_2$       & IL-2           concentration     &  ng mL$^{-1}$  \\
		$I_{12}$    & IL-12          concentration     &  ng mL$^{-1}$ \\
		$I_{10}$    & IL-10          concentration     &  ng mL$^{-1}$  \\
		$T_\beta$   & TGF-$\beta$    concentration     &  ng mL$^{-1}$  \\
		$I_\gamma$  & IFN-$\gamma$   concentration     &  ng mL$^{-1}$  \\ 
		\midrule
		$P$         & PD-1           concentration     &   nmol L$^{-1}$ \\
		$L$         & PD-L1          concentration     &  nmol L$^{-1}$ \\	
		$A$         & Anti-PD-L1      concentration     &  nmol L$^{-1}$ \\	
		$Q$         & Adjuvant concentration    & $\mu$g L$^{-1}$  \\
		$S$         & Tumor-specific antigen concentration & $\mu$g L$^{-1}$   \\
		\toprule
	\end{tabular}
\end{table}

\subsection{Cell Level: Dynamic Changes in Cell Numbers}\label{sec2.1}

\textbf{Dynamics equation of dendritic cells ($D$).} Dendritic cells, as the most critical antigen-presenting cells, transition from an immature state ($D_0$) to an activated state ($D$) upon capturing tumor-specific antigens ($S$) \cite{Wculek.NatRevImmunol.2020,Palucka.NatRevCancer.2012}. Cancer vaccines primarily consist of $S$ and adjuvants ($Q$). $Q$ drives the maturation of dendritic cells, thereby ensuring effective antigen presentation and T cell activation. The dynamics of $D$ is described as:  
\begin{equation}
	\label{Eq:D}
	\frac{\mathrm{d}D}{\mathrm{d}t}=\underbrace{\lambda_{D_0}\frac{S}{K_S+S}\left ( 1 + \lambda_{Q} \frac{Q}{K_Q+Q} \right )D_0}_{\rm{activation}} - \underbrace{d_{D}D}_{\rm{death}}.
\end{equation}
Here, $\lambda_{D_0}\frac{S}{K_S+S}\left ( 1 + \lambda_{Q} \frac{Q}{K_Q+Q} \right )D_0$ quantifies the synergistic activation of $D_0$ by $S$ and $Q$. $\lambda_{D_0}$ denotes the baseline activation rate of $D_0$. $\lambda_Q$ represents the regulatory coefficient of $Q$ on the activation process of $D_0$. $K_S$ and $K_Q$ are the half-saturation constants for $S$ and $Q$, respectively. $d_{D}$ is the death rate of $D$.

\textbf{Dynamics equation of helper T cells ($T_h$).} Na\"{i}ve CD4+ T cells ($T_{N4}$) are activated upon recognizing tumor-specific antigens presented by $D$ \cite{Wculek.NatRevImmunol.2020}. Under the regulation of IL-12 ($I_{12}$) and IFN-$\gamma$ ($I_\gamma$), $T_{N4}$ differentiates into helper T cells ($T_h$), a process suppressed by IL-10 ($I_{10}$) and TGF-$\beta$ ($T_\beta$) \cite{Trinchieri.NatRevImmunol.2003,Liew.NatRevImmunol.2002,Zhou.Immunity.2009,Zhu.Blood.2008}. Meanwhile, IL-2 ($I_2$) further promotes the proliferation of $T_h$ \cite{Spolski.NatRevImmunol.2018}. However, these processes are negatively regulated by the PD-L1 ($L$) and PD-1 ($P$) signaling pathways \cite{Okazaki.NatImmunol.2013,Morad.Cell.2022}. Therefore, anti-PD-L1 ($A$) has become an important strategy for cancer treatment. The dynamics equation for $T_h$ is described as:
\begin{equation}
	\label{Eq:T_h}
	\frac{\mathrm{d}T_{h}}{\mathrm{d}t} = \underbrace{ \left \{ J_{T_h}\cdot T_{N4} + \beta_{T_{h}} \frac{I_2}{K_{I_2}+I_2} T_{h} \right \} }_{\rm{activation\ and\ proliferation}}  \times F(P,L,A)-\underbrace{d_{T_{h}}T_{h}}_{\rm{death}},
\end{equation}
where
\begin{equation}
	\label{Eq:J_Th}
	J_{T_h}= \underbrace{\lambda_{T_h} \left ( \frac{ I_{12}}{K_{I_{12}}+I_{12}} + \frac{ I_{\gamma}}{K_{I_{\gamma}}+I_{\gamma}} \right ) }_{\rm{promotion}} \underbrace{\frac{K_{T_hI_{10}}}{K_{T_hI_{10}}+I_{10}}\frac{K_{T_hT_{\beta}}}{K_{T_hT_{\beta}}+T_{\beta}}}_{\rm{inhibition}}\underbrace{\left( \frac{D^{n}}{K^{n}_D+D^{n}} \right)}_{\rm{antigen\ presentation}},
\end{equation}
\begin{equation}
	\label{Eq:F}
	F(P,L,A)=\frac{K_{P_L}}{K_{P_L}+P_L},\ P_L=\frac{\alpha_1 L P}{1+\alpha_1 P + \alpha_2 A}.
\end{equation}
Here, $\lambda_{T_h}$, $\beta_{T_h}$, and $d_{T_h}$ represent the activation rate, proliferation rate, and death rate of $T_h$, respectively. $K_{I_2}$, $K_{I_{12}}$, and $K_{I_{\gamma}}$ denote the half-saturation constants of $I_2$, $I_{12}$, and $I_\gamma$, respectively. $K_{T_hI_{10}}$ and $K_{T_hT_\beta}$ represent the inhibitory functions of $I_{10}$ and $T_\beta$ on $T_h$ differentiation, respectively. $K_D$ denotes the half-saturation constant of $D$. $n$ represents the Hill coefficient of the antigen presentation process. Based on the previous study \cite{Li.BullMathBiol.2024}, we portrayed the action mechanisms of immune checkpoints and drugs by $F(P,L,A)$. $K_{P_L}$ represents the inhibitory effect of the PD-1-PD-L1 complex. $\alpha_1$ and $\alpha_2$ indicate the binding rates of PD-L1 to PD-1 and anti-PD-L1, respectively.

\textbf{Dynamics equation of regulatory T cells ($T_r$).} Regulatory T cells ($T_r$) are immunosuppressive CD4+ T cells that inhibit the activation of $T_h$ and $T_c$ by secreting inhibitory cytokines \cite{Zhou.Immunity.2009,Zhu.Blood.2008,Zou.NatRevImmunol.2006}. $T_\beta$ promotes the differentiation of $T_{N4}$ into $T_r$, and this process is inhibited by $I_\gamma$ \cite{Zhou.Immunity.2009,Zhu.Blood.2008,Zou.NatRevImmunol.2006}. Therefore, the dynamics equation of $T_r$ is described as:
\begin{equation}
	\label{Eq:T_r}
	\frac{\mathrm{d}T_{r}}{\mathrm{d}t} =  \underbrace{J_{T_r}\cdot T_{N4}}_{\rm{activation}} -\underbrace{d_{T_{r}}T_{r}}_{\rm{death}},
\end{equation}
where
\begin{equation}
	\label{Eq:J_Tr}
	J_{T_r}=\underbrace{\lambda_{T_r}\frac{ T_{\beta}}{K_{T_{\beta}}+T_{\beta}}}_{\rm{promotion}}\underbrace{\frac{K_{T_rI_{\gamma}}}{K_{T_rI_{\gamma}}+I_{\gamma}}}_{\rm{inhibition}}\underbrace{\left( \frac{D^{n}}{K^{n}_D+D^{n}} \right)}_{\rm{antigen\ presentation}}.
\end{equation}
Here, $\lambda_{T_r}$ and $d_{T_r}$ denote the activation rate and death rate of  $T_r$, respectively. $K_{T_\beta}$ represents the half-saturation constant of $T_\beta$. $K_{T_rI_\gamma}$ indicates the inhibitory function of $I_\gamma$ on $T_r$ differentiation. 

\textbf{Dynamics equation of cytotoxic T cells ($T_c$).} Cytotoxic T cells ($T_c$) are an important component of the adaptive immune system, derived from na\"{i}ve CD8+ T cells \cite{Philip.NatRevImmunol.2022,Giles.Immunity.2023}. $I_{12}$ promotes the differentiation of $T_{N8}$ into $T_c$, a process inhibited by $I_{10}$ and $T_\beta$ \cite{Trinchieri.NatRevImmunol.2003,Smith.Immunity.2018,Philip.NatRevImmunol.2022}. $I_2$ can promote the proliferation of $T_c$ to enhance the anti-tumor immune response \cite{Spolski.NatRevImmunol.2018}. Meanwhile, the PD-1-PD-L1 pathway inhibited the activation and proliferation of $T_c$ \cite{Okazaki.NatImmunol.2013,Morad.Cell.2022}. The mathematical framework for the $T_c$ dynamics is described as:
\begin{equation}
	\label{Eq:T_c}
	\frac{\mathrm{d}T_{c}}{\mathrm{d}t} = \underbrace{ \left \{ J_{T_c}\cdot T_{N8}+\beta_{T_{c}}\frac{I_2}{K_{I_2}+I_2}T_{c} \right \} }_{\rm{activation\ and\ proliferation}} \times F(P,L,A)-\underbrace{d_{T_{c}}T_{c}}_{\rm{death}},
\end{equation}
where
\begin{equation}
	\label{Eq:J_Tc}
	J_{T_c}=\lambda_{T_c} \underbrace{\frac{ I_{12}}{K_{I_{12}}+I_{12}} }_{\rm{promotion}} \underbrace{\frac{K_{T_cI_{10}}}{K_{T_cI_{10}}+I_{10}}\frac{K_{T_cT_{\beta}}}{K_{T_cT_{\beta}}+T_{\beta}}}_{\rm{inhibition}}\underbrace{\left( \frac{D^{n}}{K^{n}_D+D^{n}} \right)}_{\rm{antigen\ presentation}}.
\end{equation}
Here , $\lambda_{T_c}$, $\beta_{T_c}$, and $d_{T_c}$ denote the activation, proliferation and death rates of  $T_c$, respectively. $K_{T_cI_{10}}$ and $K_{T_cT_\beta}$ denote the inhibitory function of $I_{10}$ and $T_\beta$ on $T_c$ differentiation, respectively.

\textbf{Dynamics equation of tumor cells ($C$).} We assume that tumor cell ($C$) proliferation follows the Logistic equation. Based on \cite{Lai.PNAS.2018}, a bilinear form is used to describe the killing of $C$ by $T_c$ and $T_h$. Thus, the dynamics of $C$ by growth, killing, and death mechanisms are described as: 
\begin{equation}
	\label{Eq:C}
	\frac{\mathrm{d}C}{\mathrm{d}t} = \underbrace{\beta_C \left ( 1-\frac{C}{G_C} \right ) C}_{\rm{growth}} -\underbrace{ \left ( \eta_{T_c} T_c + \eta_{T_h} T_h \right )C}_{\rm{killing}}-\underbrace{ d_C C}_{\rm{death}},
\end{equation} 
where $\beta_C$, $G_C$, and $d_c$ denote the proliferation rate, carrying capacity, and death rate of $C$, respectively. $\eta_{T_c}$ and $\eta_{T_h}$ represent the killing rate of $T_c$ and $T_h$ on $C$, respectively. 

\subsection{Molecular level: Dynamic Changes in Protein Concentrations}\label{sec2.2}

\textbf{Dynamics equation of IL-2 ($I_2$).} IL-2 ($I_2$) is a pleiotropic cytokine that primarily promotes effector T cells' proliferation \cite{Spolski.NatRevImmunol.2018}. $T_h$ and $T_c$ are the main sources of $I_2$ \cite{Spolski.NatRevImmunol.2018,Propper.NatRevClinOncol.2022}. The dynamics equation of IL-2 is expressed as:
\begin{equation}
	\label{Eq:I_2}
	\tau \frac{\mathrm{d}I_2}{\mathrm{d}t} = \underbrace{\delta_{I_2T_h}\cdot T_h}_{\rm{secretion}} + \underbrace{\delta_{I_2T_c}\cdot T_c}_{\rm{secretion}} -\underbrace{d_{I_2}\cdot I_2}_{\rm{degradation}},
\end{equation}
where $\delta_{I_2T_h}$ and $\delta_{I_2Y_c}$ are the production rates of $I_2$ by $T_h$ and $T_c$, respectively. $d_ {I_2}$ represents the degradation rate of $I_2$. In the molecular dynamics, $\tau \ll 1$  represents the timescale parameter. Quasi-steady-state approximation methods are employed to integrate temporal scale differences among distinct biological processes (See Section 3.2, step (2)).

\textbf{Dynamics equation of IL-10 ($I_{10}$).} IL-10 ($I_{10}$) is a multifunctional negative regulator that plays an important role in suppressing excessive immune responses \cite{Smith.Immunity.2018}. The primary sources of $I_{10}$ include $C$ and $T_r$ \cite{Zhu.Blood.2008,Zhou.Immunity.2009,Lai.PNAS.2018}. Therefore, the dynamics equation of $I_{10}$ is described as:
\begin{equation}
	\label{Eq:I_10}
	\tau \frac{\mathrm{d}I_{10}}{\mathrm{d}t} = \underbrace{\delta_{I_{10}C}\cdot C}_{\rm{secretion}} + \underbrace{\delta_{I_{10}T_r}\cdot T_r}_{\rm{secretion}} -\underbrace{d_{I_{10}}\cdot I_{10}}_{\rm{degradation}},
\end{equation}
where $\delta_{I_{10}C}$ and $\delta_{I_{10}T_r}$ represent the production rates of $I_{10}$ by $C$ and $T_r$, respectively. $d_{I_{10}}$ represents the degradation rate of $I_{10}$.

\textbf{Dynamics equation of IL-12 ($I_{12}$).} IL-12 ($I_{12}$) is a key factor linking innate and adaptive immunity \cite{Trinchieri.NatRevImmunol.2003}. $I_{12}$ is mainly secreted by $D$ \cite{Trinchieri.NatRevImmunol.2003,Zhu.Blood.2008}. The dynamics equation for $I_{12}$ is described as:
\begin{equation}
	\label{Eq:I_12}
	\tau \frac{\mathrm{d}I_{12}}{\mathrm{d}t} = \underbrace{\delta_{I_{12}D}\cdot D}_{\rm{secretion}}  -\underbrace{d_{I_{12}}\cdot I_{12}}_{\rm{degradation}},
\end{equation}
where $\delta_{I_{12}D}$ is the production rate of $I_{12}$ by $D$. $d_{I_{12}}$ represents the degradation rate of \ $I_{12}$.

\textbf{Dynamics equation of IFN-$\gamma$ ($I_{\gamma}$).} IFN-$\gamma$ ($I_{\gamma}$) is a cytokine with anti-viral, anti-tumor, and immunomodulatory functions \cite{Boehm.AnnuRevImmunol.1997}. $I_{\gamma}$ is mainly secreted by $T_h$ and $T_c$ \cite{Boehm.AnnuRevImmunol.1997,Liew.NatRevImmunol.2002}. The dynamics equation of $I_{\gamma}$ is described as:
\begin{equation}
	\label{Eq:I_gamma}
	\tau \frac{\mathrm{d}I_\gamma}{\mathrm{d}t} = \underbrace{\delta_{I_\gamma T_h}\cdot T_h}_{\rm{secretion}} + \underbrace{\delta_{I_\gamma T_c}\cdot T_c}_{\rm{secretion}} -\underbrace{d_{I_\gamma}\cdot I_\gamma}_{\rm{degradation}},
\end{equation}
where $\delta_{I_\gamma T_h}$ and $\delta_{I_\gamma T_c}$ are the production rates of $I_\gamma$ by $T_h$ and $T_c$, respectively. $d_{I_\gamma}$ denotes the degradation rate of $I_\gamma$.

\textbf{Dynamics equation of TGF-$\beta$ ($T_{\beta}$).} TGF-$\beta$ ($T_{\beta}$) is a key immunoregulatory cytokine derived from $C$ and $T_r$ that potently suppresses $T_c$ and $T_h$ differentiation \cite{Visser.Leukemia.1999,Zhu.Blood.2008,Lai.PNAS.2018}. The dynamics equation for $T_{\beta}$ is described as:
\begin{equation}
	\label{Eq:T_beta}
	\tau \frac{\mathrm{d}T_{\beta}}{\mathrm{d}t} = \underbrace{\delta_{T_{\beta}C}\cdot C}_{\rm{secretion}} + \underbrace{\delta_{T_{\beta}T_r}\cdot T_r}_{\rm{secretion}} -\underbrace{d_{T_{\beta}}\cdot T_{\beta}}_{\rm{degradation}},
\end{equation}
where $\delta_{T_{\beta}C}$ and $\delta_{T_{\beta}T_r}$ are the production rates of $T_{\beta}$ by $C$ and $T_r$, respectively. $d_{T_{\beta}}$ denotes the degradation rate of $T_{\beta}$.

\subsection{Pharmacokinetic models of cancer vaccines and anti-PD-L1}\label{sec2.3}

In the simplified pharmacokinetic (PK) model, the concentration of the drug $X$ at time $t$ is governed by the following equation:
\begin{equation}
	\frac{\mathrm{d} X(t)}{\mathrm{d} t}=X^* \sum_{i=1}^n \delta\left(t-t_i\right)-\mu_X \cdot X(t),
\end{equation}
where $X^*$ denotes the administered dose per injection, $\mu_X$ is the elimination rate constant, and $t_{1/2} = \textrm{ln}(2)/\mu_X$ is the half-life of the drug. The term $\delta(t-t_i)$ represents the Dirac delta function, which is a generalized function satisfying $\int_{-\infty}^{\infty} \delta\left(t-t_i\right) \mathrm{d} t=1$ and $\delta\left(t-t_i\right)=0$ for all $t \neq t_i$. This function models an instantaneous drug input at time $t_i$ for $i = 1, 2, \cdots n$.

The resulting drug concentration after multiple doses is given by:
\begin{equation}
	X(t) = \sum^n_{i=1} X^* e^{-\mu_X (t-t_i)} \cdot H(t-t_i),
\end{equation}
where $e^{-\mu_X (t-t_i)}$ represents exponential clearance since the time of each injection, and $H (t-t_i) $ is the Heaviside step function, defined as $H(t-t_i) = 1$ for $t \ge t_i$ and $H(t-t_i) = 0$ for $t < t_i$. This ensures that each dose contributes only from its administration time onward.

\textbf{Dynamics equation of cancer vaccines ($S/Q$).} Cancer vaccine is an immunotherapy that uses tumor-specific antigen ($S$) to induce the production of tumor-specific T cells. Meanwhile, adjuvants ($Q$) are often added to vaccines to enhance the immune response. Under no drug treatment, the dynamics equation for the secretion of $S$ by tumor cells can be described as:
\begin{equation}
	\tau \frac{\mathrm{d}S}{\mathrm{d}t} = \underbrace{\delta_{SC} \cdot C}_{\rm{secretion}}-\underbrace{d_{S} \cdot S}_{\rm{degradation}},
\end{equation}
where $\delta_{SC}$ denotes the production rate of $S$ by $C$, $d_{S}$ denotes the degradation rate of $S$. During the rapid endogenous natural degradation process, the quasi-steady state approximation yields $S(t)=\frac{\delta_{SC}}{d_{S}}\cdot C(t)$. Therefore, the PK model of $S$ is designed as: 
\begin{equation}
	\label{Eq:S}
	S(t) = \frac{\delta_{SC}}{d_{S}} \cdot C(t) + \sum^n_{i=1} S^* e^{-\mu_S (t-t_i)} \cdot H(t-t_i),
\end{equation}
where $S^*$ and $\mu_S$ represent the administered dose and clearance rate constant, respectively. The term $\sum^n_{i=1} S^* e^{-\mu_S (t-t_i)} \cdot H(t-t_i)$ represents the relatively slow exogenous immune clearance mediated by the immune system.

Meanwhile, the PK model of $Q$ is designed as:
\begin{equation}
	\label{Eq:Q}
	Q(t) = \sum^n_{i=1} Q^* e^{-\mu_Q (t-t_i)} \cdot H(t-t_i),
\end{equation}
where $Q^*$ and $\mu_Q$ represent the administered dose and clearance rate constant, respectively.

\textbf{Dynamics equation of PD-1 ($P$), PD-L1 ($L$) and anti-PD-L1 ($A$).} PD-1 ($P$) is mainly expressed on activated T cells \cite{Okazaki.NatImmunol.2013,Morad.Cell.2022}. We assume that the expression rate of $P$ on the surface of $T_1$ and $T_8$ cells is $\rho_P$. Therefore, the concentration of $P$ is described as $P=\rho_P(T_1+T_8)$. Meanwhile, PD-L1 ($L$) is expressed on the surface of $T_1$, $T_8$ and $C$ \cite{Okazaki.NatImmunol.2013,Morad.Cell.2022}. We assume that the expression rate of $L$ on the $T_1$ and $T_8$ surfaces is $\rho_L$. Due to the high expression of $L$ on the surface of $C$, we introduce the parameter $\varepsilon_C$ to regulate it. Therefore, the concentration of $L$ is designed as $L=\rho_L\left(T_1+T_8+\varepsilon_C  C \right)$.
Anti-PD-L1 ($A$) is an immune checkpoint inhibitor that restores the anti-tumor activity of T cells by blocking the binding of PD-L1 to PD-1. The PK model for $A$ is designed as:
\begin{equation}
	\label{Eq:A}
	A(t) = \sum^n_{i=1} A^* e^{-\mu_A (t-t_i)} \cdot H(t-t_i),
\end{equation}
where $A^*$ and $\mu_A$ denote the administration dose and clearance rate constant, respectively.

\section{Methods}\label{sec3}

\subsection{Experimental data and parameters}\label{sec3.1}

The average diameter of mammalian cells is about 5 $\sim$ 20 $\mathrm{\mu m}$ \cite{Moran.Cell.2010}. We assume that the diameter of a single tumor cell is \ $d=10\ \mathrm{\mu m}$. The volume of a single tumor cell is $V_{cell}=\frac{4}{3}\pi r^3=523.6\ \mathrm{\mu m^3}$, $r=\frac{d}{2}=5\ \mu m$. Then, we converted the unit of tumor volume from $\mathrm{mm^3}$ to $\mathrm{\mu m^3}$ ($\mathrm{mm^3}=1\times10^9\ \mathrm{\mu m^3}$) in animal experiments and calculated the number of tumor cells per unit volume by the formula $N=\frac{V_{tumor}\times10^9\times\eta}{V_{cell}}$. Here, $\eta=0.64$ represents the stacking efficiency, which is based on the classical theory of stochastic close-packing to describe the phenomenon that cell stacking is not perfectly tight \cite{Torquato.PhysRevLett.2000}. Thus, the cellular density of tumor tissue measures $1.22\times 10^6\ \mathrm{cells/mm^3}$. Table \ref{Tab:Data} shows the dynamic changes in the number of tumor cells under different treatment strategies. Parameter estimation is described in the Appendix. The specific parameter values and biological significance are described in the Table \ref{Tab:Parameter}.

\begin{table}[h!]
	\tiny \centering
	\caption{Number of cells per mouse at different time points under different treatment options.}
	\label{Tab:Data}
	\renewcommand\arraystretch{0.5}
	\begin{tabular}{>{\centering\arraybackslash}p{3.9cm}|>{\centering\arraybackslash}p{2.9cm}|>{\centering\arraybackslash}p{2.9cm}|>{\centering\arraybackslash}p{2.9cm}|>{\centering\arraybackslash}p{2.9cm}}
		\toprule
		\multirow{2}{*}{ID} & \multicolumn{4}{c}{Dynamic evolution of tumor cells*} \\ 
		\cmidrule{2-5} 
		& 16 day & 22 day & 27 day & 32 day \\
		\midrule
		\multicolumn{5}{l}{\textbf{Treatment option 1: control group ($\times10^8$ cells)}} \\ 
		\midrule
		Mouse 1 & 1.4319 & 3.2589 & 8.4117 & 18.2619  \\
		Mouse 2 & 1.5331 & 4.1907 & 7.6131 & 15.2524  \\
		Mouse 3 & 2.0277 & 4.7678 & 10.4178& 19.3388  \\
		Mouse 4 & 3.8173 & 6.7097 & 12.5150 & 13.9161  \\
		Mouse 5 & 2.0309 & 5.7945 & 12.2947 & 16.3267  \\ \midrule
		
		\multicolumn{5}{l}{\textbf{Treatment option 3: vaccine group ($\times10^8$ cells)}} \\ 
		\midrule
		Mouse 6 & 1.2559 & 2.0752 & 4.9038 & 8.1544  \\
		Mouse 7 & 1.2239 & 0.8247 & 1.6045 & 3.6820  \\
		Mouse 8 & 1.1015 & 1.4284 & 3.2839 & 6.1827  \\
		Mouse 9 & 0.5161 & 0.3701 & 0.8414 & 2.5518  \\
		Mouse 10 & 1.1590 & 1.8251 & 4.3810 & 11.5794  \\
		\midrule
		
		\multicolumn{5}{l}{\textbf{Treatment option 2: anti-PD-L1 group ($\times10^8$ cells)}} \\ 
		\midrule
		Mouse 11 & 0.5177 & 0.6987 & 1.2434 & 1.9983  \\
		Mouse 12 & 0.8465 & 1.7360 & 2.7406 & 5.4598  \\
		Mouse 13 & 0.7213 & 0.8607 & 1.4762 & 2.3740  \\
		Mouse 14 & 1.1760 & 1.2912 & 1.7497 & 3.1044  \\
		Mouse 15 & 1.5324 & 2.0496 & 4.9020 & 6.3698  \\ \midrule
		
		\multicolumn{5}{l}{\textbf{Treatment option 4: vaccine + anti-PD-L1 group ($\times10^8$ cells)}} \\ 
		\midrule
		Mouse 16 & 0.4276 & 0.0001 & 0.0001 & 0.0001  \\
		Mouse 17 & 0.3914 & 0.1221 & 0.0001 & 0.0001  \\
		Mouse 18 & 0.5416 & 0.3030 & 0.2275 & 0.4207  \\
		Mouse 19 & 0.2255 & 0.0001 & 0.0001 & 0.0001  \\ 
		Mouse 20 & 1.1760 & 0.3977 & 0.1533 & 0.0001  \\
		\bottomrule
		\multicolumn{5}{@{}p{\linewidth}@{}}{\footnotesize *Note: Tumor cells were inoculated on day 0, the cancer vaccine was administered on day 12, and anti-PD-L1 treatment was delivered on days 10 and 15. Tumor volume = $0$ was processed as $1 \times 10^4$ cells (minimum detectable level). The experimental data are derived from \cite{Liu.NatCancer.2022}.} \\
	\end{tabular}
\end{table}

\begin{table}[h!]
	\tiny
	\centering
	\caption{The list of parameters and initial values of the mathematical model.}
	\label{Tab:Parameter}
	\renewcommand\arraystretch{0.5}
	\begin{tabular}{>{\centering\arraybackslash}p{2cm}|>{\centering\arraybackslash}p{6cm}|>{\centering\arraybackslash}p{1.5cm}|>{\centering\arraybackslash}p{3cm}|>{\centering\arraybackslash}p{1.5cm}}
		\toprule
		\textbf{Notation} & \textbf{Description} &  \textbf{Value} & \textbf{Units} & \textbf{References} \\ 
		\midrule
		$D$ & Initial values of dendritic cells & $3.10\times10^8$ & cells& \cite{Li.BullMathBiol.2024,Chen.MathBiosci.2022}  \\
		$T_h$ & Initial values of helper T cells & $9.50\times10^8$ & cells& \cite{Li.BullMathBiol.2024}  \\
		$T_r$ & Initial values of regulatory T cells & $6.21\times10^8$ & cells & \cite{Li.BullMathBiol.2024,Chen.MathBiosci.2022}  \\
		$T_c$ & Initial values of cytotoxic T cells & $8.04\times10^8$ & cells& \cite{Li.BullMathBiol.2024,Chen.MathBiosci.2022}   \\
		$C$ & Initial values of tumor cells & $1\times10^6$ & cells & \cite{Liu.NatCancer.2022}   \\ 
		\midrule
		$D_0$ & The number of immature DC & $1.94 \times 10^{7}$ & cells &  \cite{Chen.MathBiosci.2022} \\
		$T_{N4}$ & The number of na\"{i}ve CD4+ T cells & $3.77 \times 10^{9}$ & cells &  \cite{Rodriguez-Messan.2021.PLoSComputBiol} \\
		$T_{N8}$ & The number of na\"{i}ve CD8+ T cells & $1.61 \times 10^{9}$ & cells & \cite{Rodriguez-Messan.2021.PLoSComputBiol} \\ 
		\midrule
		$\beta_{T_h}$ & Proliferation rate of helper T cells & $0.25$ & day$^{-1}$&  \cite{Lai.SciChinaMath.2020,Friedman.BullMathBiol.2018}\\
		$\beta_{T_c}$ & Proliferation rate of cytotoxic T cells & $0.25$ & day$^{-1}$& \cite{Lai.SciChinaMath.2020,Friedman.BullMathBiol.2018} \\
		$\beta_{C}$ & Proliferation rate of tumor cells & $0.514$ &day$^{-1}$ & \cite{Pillis.CancerRes.2005,Sardar.CommunNonlinearSci.2023} \\  
		\midrule
		$\lambda_{D_0}$ & Activation rate of immature dendritic cells & $1.50$ &day$^{-1}$&  \cite{Wang.JImmunotherCancer.2021} \\
		$\lambda_{T_h}$ & Activation rate of helper T cells & $1.50$ & day$^{-1}$ &  \cite{Rodriguez-Messan.2021.PLoSComputBiol} \\ 
		$\lambda_{T_r}$ & Activation rate of  regulatory T cells & $1.50$ &day$^{-1}$ & \cite{Li.BullMathBiol.2024}\\
		$\lambda_{T_c}$ & Activation rate of  cytotoxic T cells & $16.60$ &day$^{-1}$ & \cite{Lai.SciChinaMath.2020}\\  
		\midrule
		$d_{D}$ & Death rate of mature dendritic cells & $0.01$ & day$^{-1}$ & \cite{Anbari.NPJSystBiolAppl.2024} \\
		$d_{T_h}$ & Death rate of helper T cells & $0.10$ & day$^{-1}$&  \cite{Li.BullMathBiol.2024}\\
		$d_{T_r}$ & Death rate of regulatory T cells & $0.10$ & day$^{-1}$ &  \cite{Li.BullMathBiol.2024} \\ 
		$d_{T_c}$ & Death rate of cytotoxic T cells & $0.10$ &day$^{-1}$ &  \cite{Li.BullMathBiol.2024} \\
		$d_{C}$ & Death rate of tumor cells & $0.14$ &day$^{-1}$ &  \cite{Li.BullMathBiol.2024} \\  
		\midrule
		$K_{I_{2}}$ & Half-saturation constant of IL-2 & $150$ & ng mL$^{-1}$ &  Est.   \\
		$K_{I_{12}}$ & Half-saturation constant of IL-12 & $300$ & ng mL$^{-1}$ & Est. \\
		$K_{I_{\gamma}}$ & Half-saturation constant of IFN-$\gamma$ & $80$ & ng mL$^{-1}$ & Est. \\
		$K_{T_{\beta}}$ & Half-saturation constant of TGF-$\beta$ & $0.21$ &ng mL$^{-1}$ & \cite{Li.BullMathBiol.2024} \\
		$K_{T_hI_{10}}$ & Inhibition of function of $T_h$ by IL-10 & $1.50$ & ng mL$^{-1}$ &  Est. \\
		$K_{T_hT_{\beta}}$ & Inhibition of function of $T_h$ by TGF-$\beta$ & $0.80$ & ng mL$^{-1}$ & Est \\
		$K_{T_rI_{\gamma}}$ & Inhibition of function of $T_r$ by IFN-$\gamma$ & $40$ &ng mL$^{-1}$ & Est. \\
		$K_{T_cI_{10}}$ & Inhibition of function of $T_c$ by IL-10 & $1.50$ &ng mL$^{-1}$ & Est. \\
		$K_{T_cT_{\beta}}$ & Inhibition of function of $T_c$ by TGF-$\beta$ & $0.80$ &ng mL$^{-1}$ & Est. \\ 
		\midrule
		$\delta_{I_2T_h}$ & Production rate of IL-2 by $T_h$ & $5.00 \times 10^{-7}$ & ng mL$^{-1}$ day$^{-1}$ cell$^{-1}$ & \cite{Li.BullMathBiol.2024} \\
		$\delta_{I_2T_c}$ & Production rate of IL-2 by $T_c$ & $1.00 \times 10^{-8}$ & ng mL$^{-1}$ day$^{-1}$ cell$^{-1}$ &  Est. \\
		$\delta_{I_{10}C}$ & Production rate of IL-10 by $C$ & $1.30\times10^{-10}$ & ng mL$^{-1}$ day$^{-1}$ cell$^{-1}$ &  \cite{Robertson-Tessi.JTheorBiol.2012}  \\
		$\delta_{I_{10}T_r}$ & Production rate of IL-10 by $T_r$ & $1.40\times10^{-8}$ & ng mL$^{-1}$ day$^{-1}$ cell$^{-1}$ &  \cite{Robertson-Tessi.JTheorBiol.2012} \\
		$\delta_{I_{12}D}$ & Production rate of IL-12 by $D$ & $9.00 \times 10^{-7}$ & ng mL$^{-1}$ day$^{-1}$ cell$^{-1}$ &  \cite{Li.BullMathBiol.2024} \\
		$\delta_{I_{\gamma}T_h}$ & Production rate of IFN-$\gamma$ by $T_h$ & $6.50 \times10^{-8}$ & ng mL$^{-1}$ day$^{-1}$ cell$^{-1}$ & \cite{Li.BullMathBiol.2024} \\
		$\delta_{I_{\gamma}T_c}$ & Production rate of IFN-$\gamma$ by $T_c$ & $2.50 \times10^{-7}$ & ng mL$^{-1}$ day$^{-1}$ cell$^{-1}$ &  \cite{Li.BullMathBiol.2024,Zhang.TheoryBiosci.2025} \\
		$\delta_{T_{\beta}C}$ & Production rate of TGF-$\beta$ by $C$ & $1.10 \times10^{-7}$ & ng mL$^{-1}$ day$^{-1}$ cell$^{-1}$ & \cite{Robertson-Tessi.JTheorBiol.2012} \\
		$\delta_{T_{\beta}T_r}$ & Production rate of TGF-$\beta$ by $T_r$ & $1.80 \times10^{-8}$ & ng mL$^{-1}$ day$^{-1}$ cell$^{-1}$ & \cite{Robertson-Tessi.JTheorBiol.2012} \\ 
		$\delta_{SC}$ & Production rate of $S$ by $C$ & $3.00 \times 10^{-8}$ & $\mu$g cell$^{-1}$ &  Est. \\ 
		\midrule
		$d_{I_2}$ & Degradation rate of IL-2 & $5.50$ & day$^{-1}$  & \cite{Qomlaqi.MathBiosci.2017}\\
		$d_{I_{10}}$ & Degradation rate of IL-10 & $8.32$ & day$^{-1}$  &  \cite{Lai.SciChinaMath.2020,Lai.PNAS.2018} \\
		$d_{I_{12}}$ & Degradation rate of IL-12 & $1.38$ & day$^{-1}$  &  \cite{Friedman.BullMathBiol.2018,Lai.PNAS.2018,Lai.SciChinaMath.2020} \\
		$d_{I_\gamma}$ & Degradation rate of IFN-$\gamma$ & $3.68$ & day$^{-1}$  & \cite{Liao.MathBiosci.2023} \\
		$d_{T_{\beta}}$ & Degradation rate of TGF-$\beta$ & $198$ & day$^{-1}$  & \cite{Li.BullMathBiol.2024} \\
		$d_{S}$ & Degradation rate of $S$ & $14.4$ & day$^{-1}$  & \cite{Rodriguez-Messan.2021.PLoSComputBiol} \\  
		\midrule	
		$G_{C}$ & Carrying capacity of tumor cells &  $3.00 \times10^{9}$ & cells & \cite{Ndenda.ChaosSolitonFract.2021,Qomlaqi.MathBiosci.2017} \\ 
		$K_{D}$ & Half-saturation constant of dendritic cells & $5.00 \times 10^8$ & cells & Est. \\
		$n$ & Hill coefficient of the antigen presentation & 3 & (none) & Est. \\ 
		$\rho_P$ & Expression of PD-1 in T cells & $1\times10^{-6}$ & nmol L$^{-1}$ cell$^{-1}$ & \cite{Li.BullMathBiol.2024} \\
		$\rho_L$ & Expression of PD-L1 in T cells & $2.5\times10^{-6}$ & nmol L$^{-1}$ cell$^{-1}$ & \cite{Li.BullMathBiol.2024} \\
		$\eta_{T_h}$ & killing rate of $T_h$ on $C$ & $3.00 \times 10^{-12}$ &cells$^{-1}$day$^{-1}$ & Est. \\
		$\eta_{T_c}$ & killing rate of $T_c$ on $C$ & $3.00 \times 10^{-11}$ &cells$^{-1}$day$^{-1}$ & Est. \\  $\varepsilon_C$ & Amplification coefficient of PD-L1 & $50$ & (none) & \cite{Li.BullMathBiol.2024} \\
		$K_{P_L}$ & Immune checkpoint inhibitory effect & $1000$ & nmol L$^{-1}$ & Est. \\
		$\alpha_1$ & Equilibrium constant for the PD-1-PD-L1 & $50$ & L nmol$^{-1}$ & \cite{Li.BullMathBiol.2024} \\
		$\alpha_2$ & Equilibrium constant for the PD-L1-anti-PD-L1 & $1.05\times10^{5}$ & day nmol$^{-1}$ & Est. \\ 
		$\lambda_Q$ & Regulatory coefficient of adjuvant & $25$ &(none)& Est. \\
		$K_S$ & Half-saturation constant of $S$ & $4$ & $\mu$g L$^{-1}$ & Est. \\
		$K_Q$ & Half-saturation constant of adjuvant & $10$ & $\mu$g L$^{-1}$ & Est. \\
		$\mu_S$ & Clearance rate constant of $S$ & $0.10$ & day$^{-1}$ & Est. \\
		$\mu_Q$ & Clearance rate constant of adjuvants & $0.10$ & day$^{-1}$ & Est. \\
		$\mu_A$ & Clearance rate constant of anti-PD-L1 & $0.10$ & day$^{-1}$ & Est. \\
		\bottomrule
	\end{tabular}
\end{table}

\subsection{Numerical Scheme}\label{sec3.2}

\textbf{(1) Numerical computation of slow-timescale dynamics: Euler's method.} At the cellular level, tumor-immune interactions are described by the slow-timescale dynamics model. The model characterizes temporal evolution of cell populations, where $X_i$ denotes the number of the $i$-th cell type, dynamics by the ordinary differential equation: 
\begin{equation}
	\frac{\mathrm{d}X_i }{\mathrm{d}t} = F_i(\mathbf{X},\mathbf{Y};\Theta).
\end{equation}
Here, $F_i(\mathbf{X},\mathbf{Y};\Theta )$ encodes multi-scale interaction mechanisms, with $\Theta$ representing the parameter set. The slow-timescale variables $\mathbf{X}=(D_0,D,T_{N4},T_h,T_r,T_{N8},T_c,C)$ track cellular dynamics (proliferation, differentiation, apoptosis, and intercellular interactions), while fast-timescale variables $\mathbf{Y}=(I_2,I_{10},I_{12},I_\gamma,T_\beta,S,Q,A)$ model cytokine and drug dynamics. The model was numerically solved by Euler's method.

\textbf{(2) Numerical computation of fast-timescale dynamics: Quasi-steady-state approximation.} At the molecular level, cytokine and drug dynamics are governed by fast-timescale dynamics, characterizing instantaneous concentration changes. Let $Y_i$ denote the $i$-th fast-timescale variate, described by:
\begin{equation}
	\tau \frac{\mathrm{d}Y_i}{\mathrm{d}t} = I_i(\mathbf{X};\Theta) -D_i(\mathbf{Y};\Theta)\ (\tau\ll 1).
\end{equation}
where $I_i(\mathbf{X};\Theta)=\sum_{j=1}^{n}\delta_{Y_iX_j}\cdot X_j$ represents the production rate of cytokine $Y_i$. $D_i(\mathbf{Y};\Theta)=d_i\cdot Y_i$ indicates the degradation rate of cytokine $Y_i$. $\tau$ is a small parameter reflecting rapid equilibration. Under the quasi-steady-state assumption ($\tau \to 0$), the dynamics reduce to:
\begin{equation}
	Y_i = \frac{1}{d_i} \sum_{j=1}^n \delta_{Y_iX_j} \cdot X_j.
\end{equation}

\textbf{(3) Biological implications of multiscale coupling.} The multiscale dynamical model dissects tumor-immune interactions into two distinct tiers: molecular-level (fast-timescale) and cellular-level (slow-timescale) processes. The fast-timescale variables ($\mathbf{Y}$) couple with slow-timescale variables ($\mathbf{X}$) through quasi-steady-state constraints, while $\mathbf{X}$ reciprocally modulates $\mathbf{Y}$ by dynamics mechanisms. This multi-scale coupling mechanism comprehensively reflects the dynamic characteristics of tumor-immune interactions and reveals the multilevel regulatory capacity of the immune system. The PK dynamics of antigen ($S$), adjuvant ($Q$), and anti-PD-L1 ($A$) operate on a timescale of hours to days due to processes such as absorption, distribution, and clearance, which are significantly slower than cytokine dynamics. Therefore, they are incorporated into the tumor-immune interactions through explicit expressions of algebraic equations.

\subsection{Virtual sample generation based on approximate Bayesian computation}\label{sec3.3}

To quantitatively characterize the dynamic evolution of tumor heterogeneity under different treatment strategies, a statistical inference framework based on approximate Bayesian computation (ABC) was developed in this study \cite{Toni.JRSocInterface.2009,MacLean.JRSocInterface.2013,Kypraios.MathBiosci.2017}. The framework contains the following three core modules. 

\begin{itemize}
	
	\item \textbf{(1) Prior distribution.} We define a multidimensional parameter set $\theta = \{ G_C, \beta_C, \alpha_2, \lambda_Q, K_Q \}$ to characterize the heterogeneous response to tumor treatment. Here, $G_C$ and \ $\beta_C$ denote the carrying capacity and basic proliferation rate of the tumor, respectively, which together form the basis of the intrinsic heterogeneity of tumor growth. As a key PK parameter of anti-PD-L1 therapy, the value range of $\alpha_2$ directly reflects the differences in tumor response to immune checkpoint inhibitors. $\lambda_Q$ and $K_Q$ denote the regulatory coefficients and half-saturation constants of the adjuvants, respectively, which together determine the heterogeneity of the therapeutic effects of cancer vaccines. During the parameter initialization phase, for each parameter $\theta_p$ in the set $\theta$, we perform uniform random sampling within its physiologically feasible interval $\Omega_p = (a_p, b_p)$, i.e.
	\begin{equation}
		\theta_p \sim \mathcal{U}(a_p, b_p),\ \forall\ \theta_p \in \theta.
	\end{equation}
	
	\item \textbf{(2) ABC method.} We employ ABC method to estimate the posterior distribution of parameters. Let $\theta$ denote the set of model parameters to be estimated. Given the prior distribution $\pi(\theta)$, the objective is to approximate the posterior distribution:
	\begin{equation}
		\label{Eq:P}
		P(\theta|x_0) \propto \int K_{\epsilon}(\rho(x_*,x_0),\mathbb{I}(\rho,\rho_0)) \cdot F(x_*|\theta ) \cdot \pi(\theta) d{x_*}.
	\end{equation}
	Here, $x_0$ represents the observed empirical data, $x_*$ denotes the generated simulated data. $F(x_*|\theta )$ represents the model likelihood of generating simulated data $x_*$ given the parameters $\theta$. The adaptive Gaussian kernel function with a cutoff, $K_{\epsilon}(\rho(x_*,x_0),\mathbb{I}(\rho,\rho_0))$, is defined as:
	\begin{equation}
		K_{\epsilon}(\rho(x_*,x_0),\mathbb{I}(\rho,\rho_0))=\textrm{exp}\left ( -\frac{\rho(x_*,x_0)^2}{2 \epsilon^2}  \right ) \times \mathbb{I}(\rho,\rho_0).
	\end{equation}
	Here, $\epsilon = 0.05$ denotes the bandwidth parameter, controlling the decay rate of the weights. The indicator function $\mathbb{I}(\rho,\rho_0)$ is defined as:
	\begin{equation}
		\mathbb{I}(\rho,\rho_0) =\begin{cases} 1, & \rho(x_*,x_0) \le \rho_0 \\ 0, &\rho(x_*,x_0) > \rho_0 \end{cases}.
	\end{equation}
	Here, $\rho_0 = 0.8$ denotes the cutoff threshold. The distance function $\rho(x_*,x_0)$, which measures the discrepancy between simulated data $x_*$ and observed data $x_0$, is defined as:
	\begin{equation}
		\rho(x_*,x_0) = 1 - \mathcal{R}^2,
	\end{equation}
	where $\mathcal{R}^2 = \max_{i}(R_i^2)$ denotes the best fit among all experimental samples. The coefficient of determination ($R_i$) for the $i$-th experimental sample is given by:
	\begin{equation}
		R_i^2 = 1 - \frac{\sum_{j=1}^{n}(x_0^{i,j}-x_*^{i,j})^2}{\sum_{j=1}^{n}(x_0^{i,j}-\bar{x}_0^i)^2}.
	\end{equation}
	Here, $\bar{x}_0^i=\frac{1}{n}\sum_{j=1}^n x_0^{ij}$ denotes the mean of the observed data points in the $i$-th sample, $x_0^{i,j}$ represents the $j$-th observed data point in the $i$-th sample, and $x_*^{i,j}$ corresponds to the $j$-th simulated data point in the $i$-th sample. Since $R^2 \in [-\infty,1)$, it follows that $\rho \in [0,+\infty)$. A smaller $\rho$ indicates better agreement between the simulated and observed data.
	
	\item \textbf{(3) Posterior Distribution.} Through the ABC process, we obtain a set of discrete samples $\theta = \{ \theta_1, \theta_2, \cdots , \theta_m \}$ with associated weights $\omega = \{ \omega_1, \omega_2, \cdots, \omega_m \}$. The discretized posterior distribution of parameters can thus be represented as:
	\begin{equation}
		P(\theta|x_0) \approx \sum^m_{k=1} \omega^*_k \cdot \delta(\theta - \theta_k),
	\end{equation}
	where $\omega^*_k = \omega_k / \sum^m_{k=1} \omega_k$ denotes the normalized weight, $\theta_k$ represents the parameters sampled from the prior distribution $\pi(\theta)$ in the $k$-th iteration, and $m$ indicates the number of effective parameter draws. The weight $\omega_k$ corresponds to parameter $\theta_k$ and is obtained from the kernel function $K_{\epsilon}(\rho(x_*,x_0),\mathbb{I}(\rho,\rho_0))$ computed in the $k$-th iteration. Here, $\delta(\theta - \theta_k)$ is the Kronecker delta function, which equals 1 where $\theta = \theta_k$ and 0 otherwise. Moreover, we can convert this discrete probability distribution into a continuous one through Gaussian kernel density estimation:
	\begin{equation}
		f(\theta | x_0) = \sum^m_{k=1} \omega^*_k \cdot \left( \frac{1}{h \sqrt{2\pi}} \textrm{exp}\left(-\frac{\left(\theta-\theta_k\right)^2}{2h^2} \right) \right),
	\end{equation}
	where $h$ represents the smoothing parameter.
\end{itemize}

\section{Results}\label{sec4}

\begin{figure}[h!]
	\centering
	\includegraphics[width=16cm]{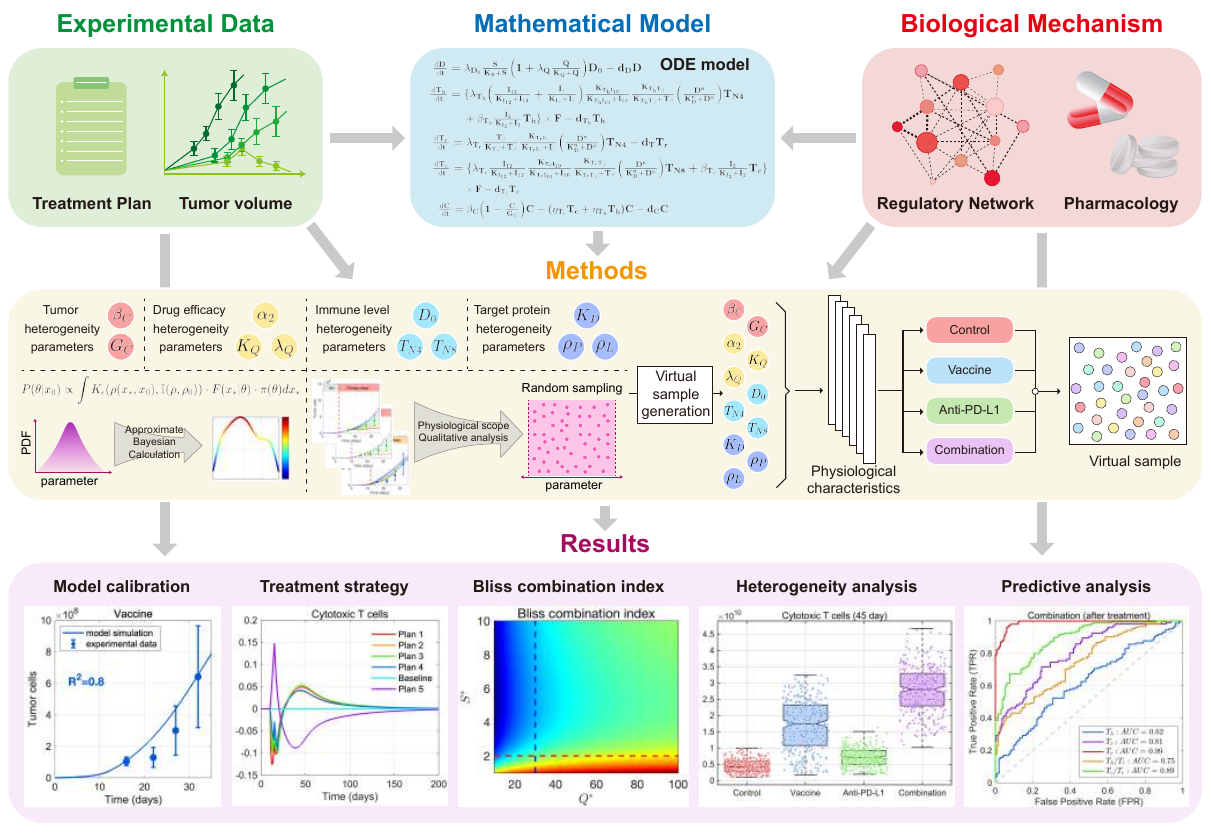}
	\caption{\textbf{Research framework overview.} This study consists of five main components. (1) Experimental data: Primarily including treatment regimens, temporal changes in tumor volume, and generated virtual sample cohort data. (2) Biological mechanisms: Mainly encompassing regulatory mechanisms of cell-cell interactions, cytokine networks, and PD principles. (3) Mathematical modeling: Integrating cell dynamics, cytokine dynamics, and PK models. (4) Methodology: Comprising approximate Bayesian computation (ABC), qualitative and quantitative parameter analysis, and virtual patient generation. (5) Results: Model-experiment fitting, therapeutic efficacy evaluation, Bliss combination index, immune cell heterogeneity distributions, and ROC analysis.}
	\label{Fig2}
\end{figure}

In this study, we integrate experimental data with biological mechanisms to construct a multiscale tumor-immune dynamic model, aiming to systematically analyze the effects of immune checkpoint blockade and cancer vaccines on tumor dynamics and the evolution of immune cell heterogeneity. The overall research design is illustrated in Fig. \ref{Fig2}.

\subsection{Tumor heterogeneity modeling and virtual cohort analysis}\label{sec4.1}

\begin{figure}[h!]
	\centering
	\includegraphics[width=16cm]{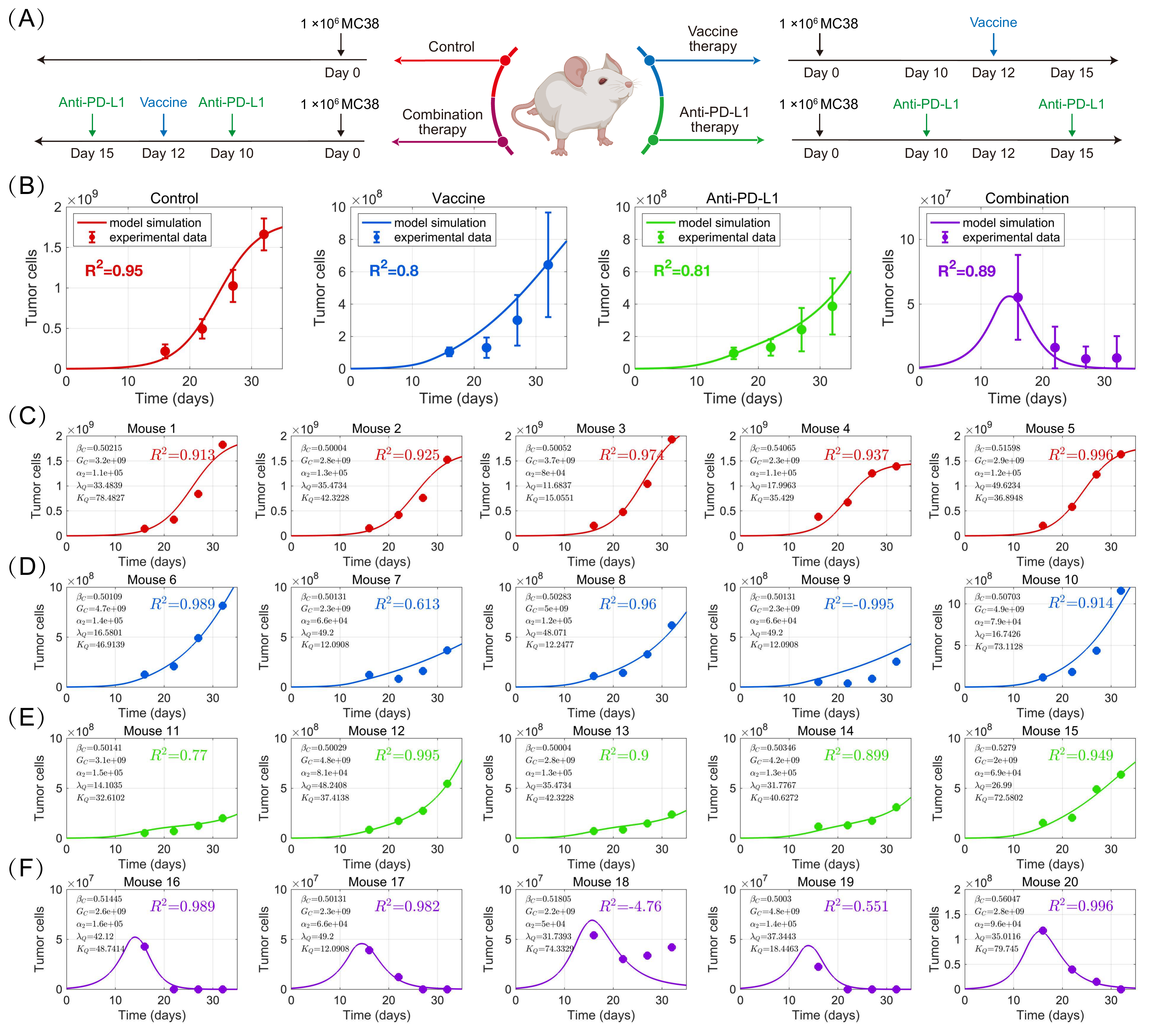}
	\caption{\textbf{Tumor evolution dynamics and model validation under different treatment strategies.} (A) Schematic of experimental design. (B) Experimental measurements (scatter points) versus model-predicted curves (solid lines) of tumor volume changes across treatment strategies. (C) $\sim$ (F) Individualized treatment response analysis: Tumor growth dynamics for 20 mice in control (C), cancer vaccine (D), anti-PD-L1 monotherapy (E), and combination therapy (F) groups. Scatter points represent experimental data derived from \cite{Liu.NatCancer.2022}, while curves show numerical results from the tumor heterogeneity model. Parameter values corresponding to the tumor heterogeneity modeling framework are indicated in the upper-left text. Panel (B) corresponds to the baseline parameters in Table \ref{Tab:Parameter}. Panels (C)-(F) display the best-fit results for 1,000 virtual patients based on approximate Bayesian parameter selection.}
	\label{Fig3}
\end{figure}

To investigate the anti-tumor effects of cancer vaccines and immune checkpoint inhibitors, we utilized tumor progression data from C57BL/6J mice bearing MC38 colorectal tumors \cite{Liu.NatCancer.2022}. The study included five mice per group: control, cancer vaccine, anti-PD-L1, and combination therapy (Fig. \ref{Fig3}A). To evaluate tumor dynamics under different treatments, we assessed model fit using the coefficient of determination ($R^2$). Computational results demonstrated that baseline parameters effectively captured tumor evolution across all groups: control ($R^2 = 0.95$), cancer vaccine ($R^2 = 0.80$), anti-PD-L1 ($R^2 = 0.81$), and combination therapy ($R^2 = 0.89$) (Fig. \ref{Fig3}B). Both monotherapies significantly suppressed tumor growth, with anti-PD-L1 exhibiting superior efficacy compared to the vaccine. Notably, the combination therapy enhanced tumor control, resulting in complete regression in some cases.

\begin{figure}[h!]
	\centering
	\includegraphics[width=16cm]{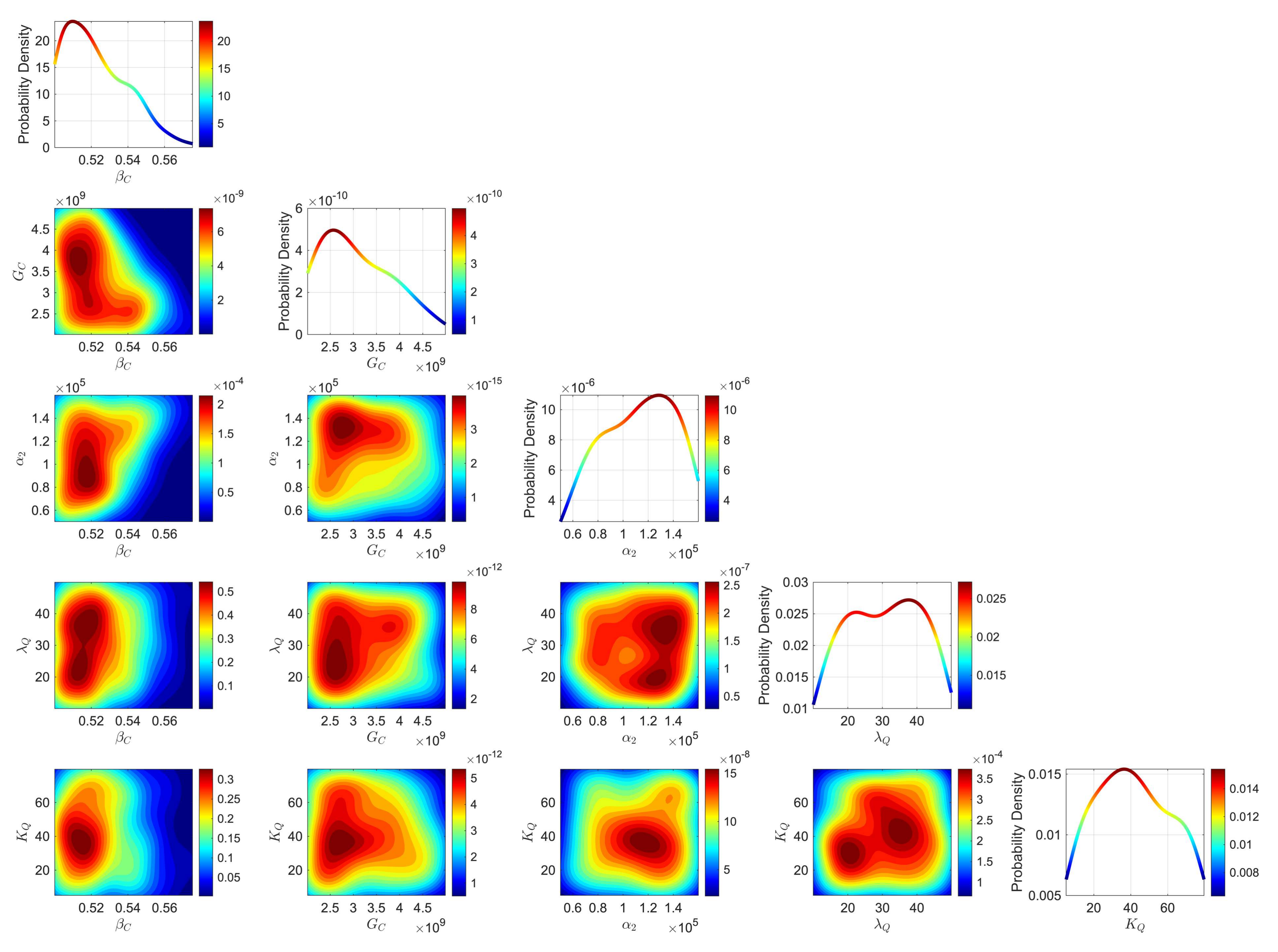}
	\caption{\textbf{Parameter posterior distribution based on approximate Bayesian calculation.} The diagonal elements represent the marginal posterior distributions calculated via weighted Gaussian kernel density estimation. The lower triangular region displays two-dimensional contour projections of different parameter combinations, with dark red areas corresponding to high probability density intervals.}
	\label{Fig4}
\end{figure}

To elucidate individualized treatment response heterogeneity, we developed a tumor heterogeneity modeling framework based on the ABC method (see Method \ref{sec3.3}). We constructed prior distributions by randomly sampling tumor heterogeneity parameters ($\beta_C \in [0.5, 0.6]$, $G_C \in [2\times 10^9, 5\times 10^9]$) and drug response heterogeneity parameters ($\alpha_2 \in [5\times 10^4, 1.6\times10^5]$, $\lambda_Q \in [10 , 50]$, $K_Q \in [5, 80]$). Based on predefined parameter distributions, we generated an initial cohort of 1000 virtual mice and simulated the immune-tumor dynamics under four different treatment strategies. Subsequently, we applied the ABC method to identify parameter sets that met a predefined goodness-of-fit threshold (Fig. \ref{Fig4}). Notably, responses such as those in Mice 9 and 18 were not fully captured by these parameter sets. Therefore, we selected parameters that best matched these atypical phenotypes from the full cohort to investigate their underlying immune mechanisms.

Individual-level fitting results demonstrated strong concordance between experimental tumor growth curves and model simulations across treatment groups (Fig. \ref{Fig3}C-F). Specifically:
\begin{itemize}	
	\item \textbf{Control group} exhibited $R^2$ values of 0.913, 0.925, 0.974, 0.937, and 0.996 (Fig. \ref{Fig3}C).
	\item \textbf{Cancer vaccine group} showed $R^2$ values of 0.989, 0.613, 0.960, -0.995, and 0.914 (Fig. \ref{Fig3}D).
	\item \textbf{Anti-PD-L1 group} demonstrated $R^2$ values of 0.770, 0.995, 0.900, 0.899, and 0.949 (Fig. \ref{Fig3}E).
	\item \textbf{Combination therapy group} displayed $R^2$ values of 0.989, 0.982, -4.760, 0.551, and 0.996 (Fig. \ref{Fig3}F).
\end{itemize}
Notably, suboptimal fitting performance ($R^2 < 0$) was observed for mouse 9 in the vaccine group and mouse 18 in the combination group. These outliers may reflect experimental variability or unmodeled biological factors influencing treatment response.

\subsection{Parameter analysis and mechanism resolution of tumor-immune regulatory networks}\label{sec4.2}

\begin{figure}[h!]
	\centering
	\includegraphics[width=16cm]{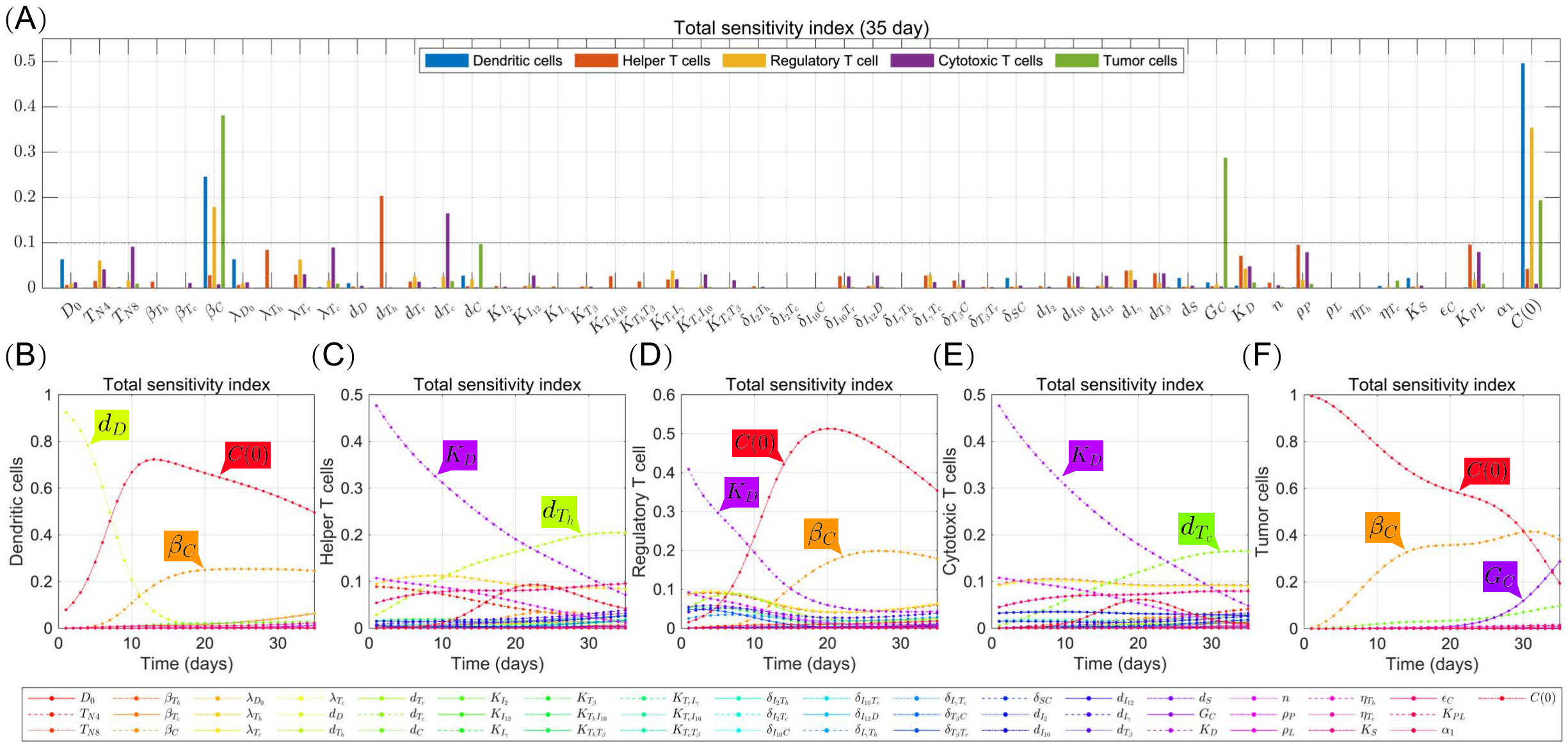}
	\caption{\textbf{Global sensitivity analysis of the baseline system without treatment.} (A) Sensitivity of system variables to model parameters was assessed using the Sobol method. Results are shown for (B) dendritic cells, (C) helper T cells, (D) regulatory T cells, (E) cytotoxic T cells, and (F) tumor cells.}
	\label{Fig5}
\end{figure}

To evaluate the influence of model parameters on output variables, we employed a global sensitivity analysis using the Sobol method \cite{Sobol.MathComputSimulat.2001}. We performed random sampling for all parameters in the untreated baseline system within a range of $\pm10\%$ and for the initial tumor volume $C(0)\in[1 \times 10^5, 1\times 10^7]$. The global sensitivity analysis results are presented in Fig. \ref{Fig5}. The numerical results showed that the population of tumor cells ($C$) was primarily influenced by the tumor proliferation rate ($\beta_C$) and initial tumor size ($C(0)$); the population of helper T cells ($T_h$) was most sensitive to its death rate ($d_{T_h}$); the regulatory T cell population ($T_r$) was modulated by $\beta_C$ and $C(0)$; and the cytotoxic T cell population ($T_c$) was predominantly controlled by its death rate ($d_{T_c}$) (Fig. \ref{Fig5}A).

Notably, the sensitivity of $C(0)$ to both $D$ and $T_r$ exhibited a rapid initial increase followed by a gradual decline over time, whereas its sensitivity to $C$ continued to decrease (Fig. \ref{Fig5}B, D, and F). This suggests that the initial tumor burden exerts a strong transient regulatory effect on immune cell dynamics in the early stages, though its direct influence diminishes over time. Simultaneously, the dependence of the tumor on its initial conditions gradually weakened, reflecting the progressive establishment of system intrinsic regulatory mechanisms. The sensitivity of $\beta_C$ to $D$, $T_r$, and $C$ increased gradually before stabilizing with minor fluctuations (Fig. \ref{Fig5}B, D, and F). This indicates that during the early stages of system evolution, the impact of tumor proliferation rates on immune cells and tumor populations gradually increases and ultimately reaches a stable state. Furthermore, we observed that the sensitivity of $d_D$ to $D$ decreased over time (Fig. \ref{Fig5}B). In contrast, the sensitivity of $d_{T_h}$ to $T_h$ and that of $d_{T_c}$ to $T_c$ increased gradually (Fig. \ref{Fig5}C and E). This result suggests that the abundances of helper T cells and cytotoxic T cells are more directly regulated by their respective death rates, and that these parameters play an increasingly important role in later immune responses. The sensitivity of $K_D$ to T cell populations declined over time (Fig. \ref{Fig5}C, D, and E), indicating a diminishing role of the half-saturation constant of dendritic cells in regulating T cell activation or proliferation, and implying that other mechanisms may take precedence in later stages. Lastly, the sensitivity of $G_C$ to $C$ increased progressively (Fig. \ref{Fig5}F), suggesting that environmental factors such as resource competition and spatial constraints become key mechanisms controlling tumor size during later phases of system evolution.

\begin{figure}[h!]
	\centering
	\includegraphics[width=16cm]{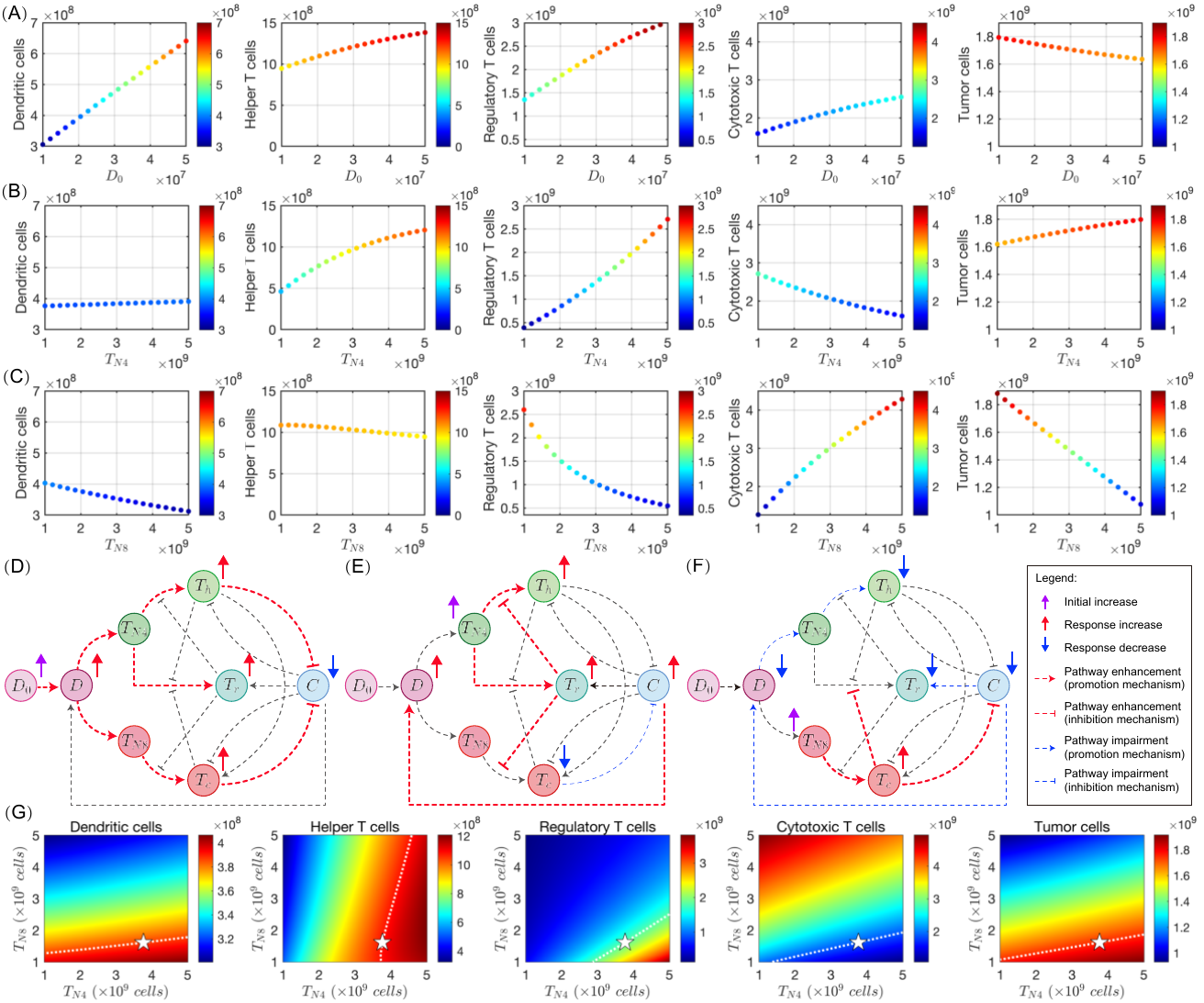}
	\caption{\textbf{Quantitative analysis of tumor-immune system regulation by immune heterogeneity parameters.} (A) $\sim$ (C) Dynamic effects of immune heterogeneity parameters ($D_0$, $T_{N4}$, and $T_{N8}$) on immune cells ($D$, $T_h$, $T_r$, $T_c$) and tumor burden ($C$).  (D) $\sim$ (F) The dynamic regulation of tumor-immune networks is mediated by increasing levels of $D_0$, $T_{N4}$, and $T_{N8}$. Purple, red, and blue arrows represent the initial increase effect, response increase, and response decrease, respectively. Black dashed lines depict baseline regulatory relationships in the tumor-immune system, while red and blue dashed lines signify strengthened and weakened interactions, respectively. (G) The two-parameter phase diagrams of $T_{N4}$ and $T_{N8}$. The white pentagram indicates the model's baseline parameter set, with the white dashed contour representing response isopleths equivalent to baseline outputs. The colorbar represents the change in cell numbers. The numerical simulation is selected from the baseline system without drug treatment. For other parameters, see Table \ref{Tab:Parameter}.}
	\label{Fig6}
\end{figure}

To investigate the impact of immune heterogeneity, we performed parameter sweeps across key immune cell populations: immature dendritic cells ($D_0 \in [1\times10^7, 5\times10^7]$), na\"{i}ve CD4+ T cells ($T_{N4} \in [1\times10^9, 5\times10^9]$), and na\"{i}ve CD8+ T cells ($T_{N8} \in [1\times10^9, 5\times10^9]$). Each parameter was sampled at 20 equidistant points, and the model outputs were simulated at day 35 (Fig. \ref{Fig6}A–C). Numerical simulations revealed that high levels of $D_0$ promote $D$ activation, which enhances adaptive immunity (elevating $T_h$, $T_r$, and $T_c$) and suppresses tumor growth (reducing $C$) (Fig. \ref{Fig6}A and D). Meanwhile, $T_{N4}$ increased the abundance of $D$, $T_h$, $T_r$, and $C$ while suppressing $T_c$ (Fig. \ref{Fig6}B). This occurs because $T_{N4}$ enhances $T_h$ and $T_r$ activation, but the resulting $T_r$ inhibits $T_c$ differentiation, thereby promoting $C$ growth, which in turn enhances $D$ maturation (Fig. \ref{Fig6}E). The perturbation results for $T_{N8}$ exhibit a completely opposite trend (Fig. \ref{Fig6}C and F). Collectively, in the high $T_{N4}$ \& low $T_{N8}$ pattern, $D$, $T_h$, $T_r$, and $C$ accelerate activation (Fig. \ref{Fig6}G). In the low $T_{N4}$ \& high $T_{N8}$ pattern, $T_c$ accelerates activation and inhibits $C$ expansion (Fig. \ref{Fig6}G).

\subsection{Efficacy analysis of immune checkpoint expression levels and anti-PD-L1 dosing plans}

\begin{figure}[h!]
	\centering
	\includegraphics[width=16cm]{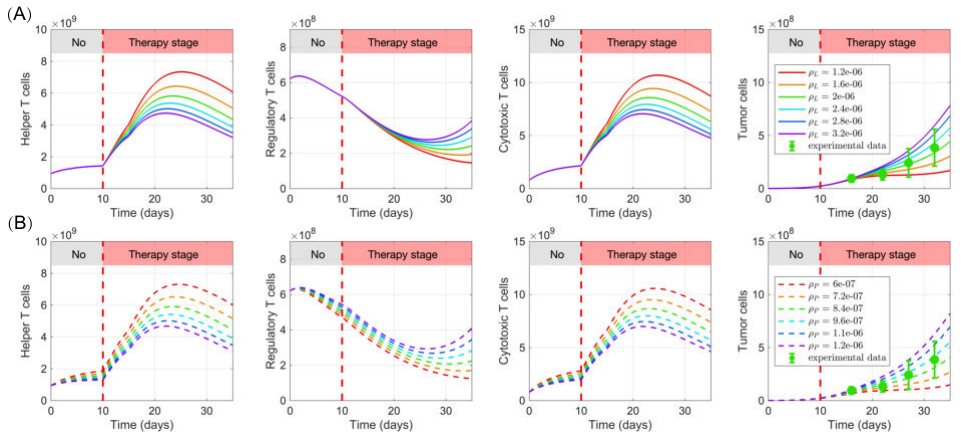}
	\caption{\textbf{Effect of differential expression of PD-L1 (A) and PD-1 (B) on tumor-immune system dynamics under anti-PD-L1 therapy.} All parameters except for the expression rates of PD-L1 ($\rho_L$) and PD-1 ($\rho_P$) are listed in Table \ref{Tab:Parameter}. The treatment option follows the standard anti-PD-L1 therapy (administering the anti-PD-L1 on days 10 and 15).}
	\label{Fig7}
\end{figure}

To systematically evaluate the response of anti-PD-L1 treatment to the expression rates of PD-L1 ($\rho_L$) and PD-1 ($\rho_P$), we conducted qualitative analyses of $\rho_L$ and $\rho_P$. Using the baseline PD-L1 expression level ($\rho_L = 2.5 \times 10^{-6}$) as reference, we tested six equally spaced values ($1.2$–$3.2 \times 10^{-6}$) and tracked tumor and T cell dynamics. Computational results demonstrated that upregulated PD-L1 expression significantly promoted the expansion of regulatory T cells ($T_r$) and tumor cells ($C$) while inhibiting the activation of helper T cells ($T_h$) and cytotoxic T cells ($T_c$) (Fig. \ref{Fig7}A). Similarly, using the baseline PD-1 expression level ($\rho_P = 1\times 10^{-6}$) as reference, we established six expression gradients ($6 \times 10^{-7} - 1.2 \times 10^{-6}$). Results showed that increasing PD-1 expression rates significantly elevated $T_r$ and $C$ populations while gradually decreasing $T_h$ and $T_c$ numbers (Fig. \ref{Fig7}B). Notably, PD-L1 mainly influences immunodynamics during treatment, whereas high PD-1 expression persists to modulate the tumor microenvironment even in the untreated phase. 

\begin{figure}[h!]
	\centering
	\includegraphics[width=16cm]{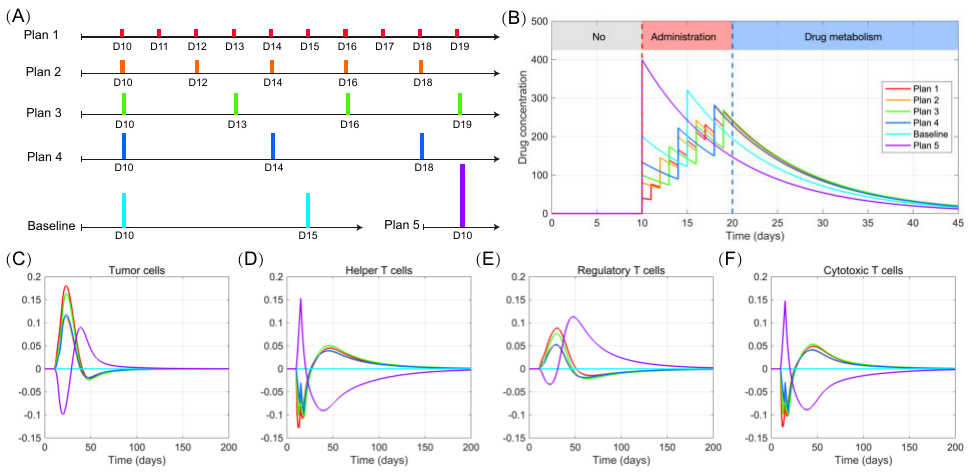}
	\caption{\textbf{Pharmacodynamic evaluation of anti-PD-L1 in different plans.} (A) Administration time and dosage design for different plans. Plan 1: 1-day interval therapy; Plan 2: 2-day interval therapy; Plan 3: 3-day interval therapy; Plan 4: 4-day interval therapy; Plan 5: Maximum dose therapy. All five plans maintain the same total dosage as the baseline plan over a 10-19 day treatment period. (B) PK curves under different plans. (C) $\sim$ (F) Relative changes in $C$, $T_h$, $T_r$, and $T_c$ under different plans relative to the baseline plan (cyan curve). Parameter values are given in Table \ref{Tab:Parameter}.}
	\label{Fig8}
\end{figure}

To compare the treatment efficacy of different anti-PD-L1 dosing regimens, we tested five plans (Fig. \ref{Fig8}A). Simulations showed that maximum dose therapy (Plan 5) achieved higher peak concentrations but also faster clearance, leading to pronounced troughs during metabolism (Fig. \ref{Fig8}B). Multiple low-dose plans (Plans 1–4) maintained more sustained drug levels compared to the baseline (Fig. \ref{Fig8}B).

To quantitatively compare the effects of different plans on the tumor-immune system, we introduced an index:
\begin{equation}
	\Phi_{ij} = \frac{R^i_j-R_j^*}{R_j^*}, j = \{ C, T_h, T_r, T_c \},
\end{equation}
where $R_j^*$ represents the output of model variable $j$ under baseline plan, $R^i_j$ represents the output of model variable $j$ under plan $i$. Initially failed to control tumor ($C$) proliferation compared to the baseline due to lower drug exposure (days 10–40), they ultimately achieved sustained growth inhibition (Fig. \ref{Fig8}C). In contrast, the maximum dose therapy (Plan 5) showed potent early inhibition (days 10–30) but resulted in rapid tumor rebound ($C$) following drug clearance (Fig. \ref{Fig8}C). Furthermore, while the multiple low-dose plans temporarily suppressed the infiltration of $T_h$ and $T_c$ during administration, they downregulated $T_r$ levels in the later phase (Fig. \ref{Fig8}D-F). Conversely, Plan 5 induced a rapid expansion of $T_h$ and $T_c$ during treatment (Fig. \ref{Fig8}D and F), yet was followed by a gradual increase in $T_r$ infiltration after drug metabolism (Fig. \ref{Fig8}E).

These results suggest that the sustained drug exposure produced by multiple low-dose plans is more conducive to maintaining long-term immune activation. Conversely, the maximum dose therapy may enhance the short-term efficacy of the drug, but may affect the long-term efficacy due to the subsequent expansion of immunosuppressive cells. 

\subsection{Efficacy analysis of antigen activation levels and cancer vaccines}

\begin{figure}[h!]
	\centering
	\includegraphics[width=16cm]{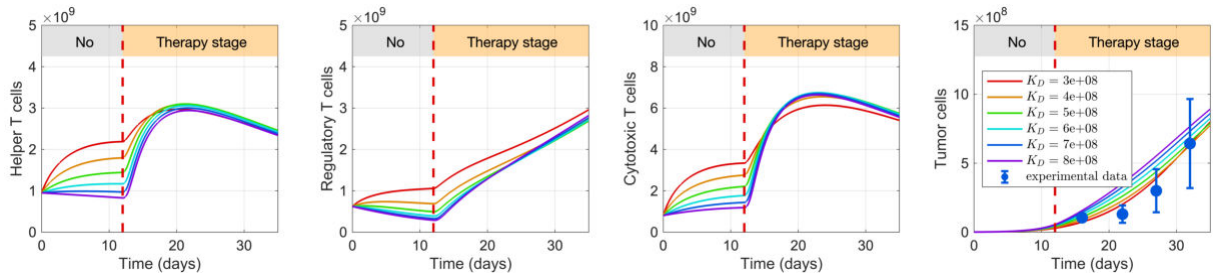}
	\caption{\textbf{Effects of cancer vaccine therapy on the evolution of tumor-immune system dynamics at different antigen presentation levels.} All parameters except for the half-saturation constant of dendritic cells ($K_D$) are listed in Table \ref{Tab:Parameter}. The treatment option follows the standard cancer vaccine therapy (administering the vaccine and adjuvant on day 12).}
	\label{Fig9}
\end{figure}

To analyze the effect of different levels of antigen presentation on the evolution of the tumor-immune system dynamics, we performed a qualitative analysis of the half-saturation constant of dendritic cells ($K_D$). Based on the baseline value of $K_D = 5 \times 10^8$, we set six half-saturation constant levels ($3-8 \times 10^8$), analyzing the dynamic evolution of tumors and key T cell subsets. $K_D$ reflects the efficacy of antigen presentation. A smaller $K_D$ value indicates that fewer dendritic cells (presenting more antigen per cell) are required to activate T cells. The results showed that as $K_D$ increases, the activation of helper T cells ($T_h$), regulatory T cells ($T_r$), and cytotoxic T cells ($T_c$) is suppressed, while tumor cells ($C$) expand rapidly. This is consistent with the mechanism of tumor immune escape, i.e., when antigen presentation is insufficient, the anti-tumor immune response cannot be activated effectively. In addition, higher $K_D$ showed better immune dynamics after receiving the cancer vaccine, especially the rapid amplification of $T_h$ and $T_c$ (Fig. \ref{Fig9}). 

\begin{figure}[h!]
	\centering
	\includegraphics[width=16cm]{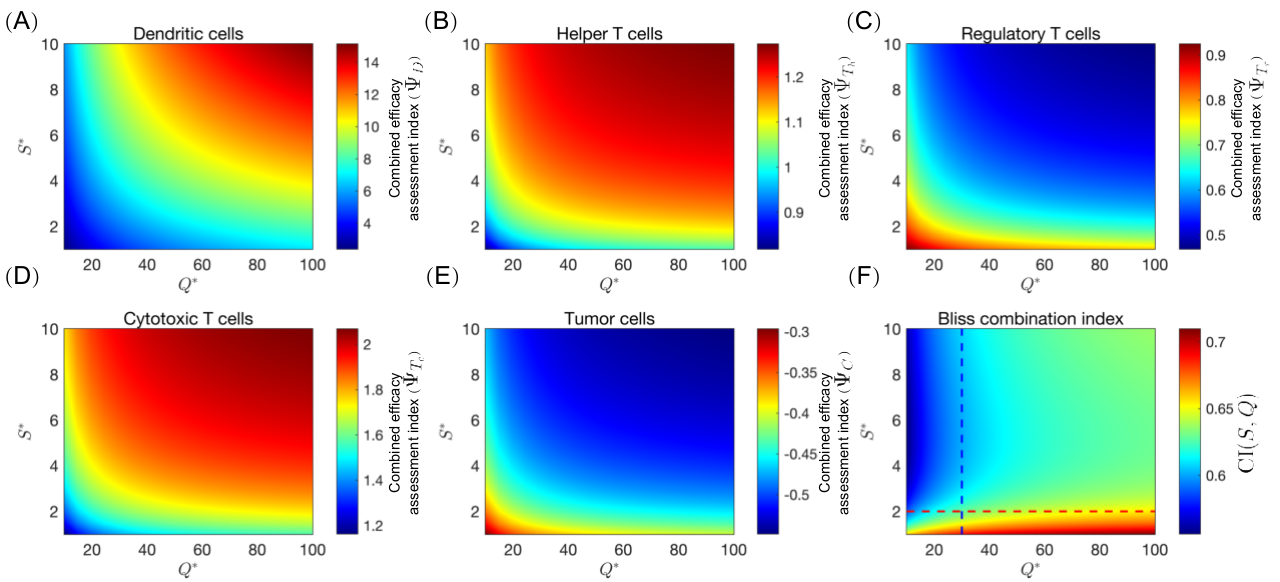}
	\caption{\textbf{Dynamic regulation of the tumor microenvironment by antigen-adjuvant dose combinations.} (A) $\sim$ (E) Heatmap showing the relative change rate of $D$, $T_h$, $T_r$, $T_c$, and $C$ at different dose combinations. (F) Assessment of synergistic effects of antigens and adjuvants based on the Bliss combination index.}
	\label{Fig10}
\end{figure}

To quantitatively evaluate the effect of dose combinations of antigen ($S^*$) and adjuvant ($Q^*$) on tumor-immune dynamics, we established a combined efficacy assessment index:
\begin{equation}
	\Psi _{i}(S^*,Q^*) = \frac{R_i(S^*,Q^*)-R_i(0,0)}{R_i(0,0)}, i = \{ D,T_h, T_r, T_c, C \},
\end{equation}
where $R_i(0,0)$ represents the output of the model variable $i$ on day 35 in the control group, and $R_i(S^*,Q^*)$ represents the output of the model variable $i$ on day 35 at the administered dose $(S^*\in [1,10 \mathrm{\mu g}],Q^*\in [10,100 \mathrm{\mu g}])$. $\Psi _{i}>0$ indicates enhancement effects, while $\Psi _{i}<0$ indicates suppression effects. Calculated results showed that with increasing doses of antigen and adjuvant, $D$, $T_h$, and $T_c$ were significantly activated (Fig. \ref{Fig10}A, B, and D). This indicated that $D$, $T_h$, and $T_c$ were effectively activated under the action of high doses of antigen and adjuvant. On the contrary, $D$ and $T_h$ were suppressed (Fig. \ref{Fig10}C and E), which means that the tumor growth was controlled.

To further evaluate the synergistic effect of antigens and adjuvants in cancer vaccines on the tumors, we introduced the Bliss combination index \cite{Foucquier.PharmacolResPerspect.2015,Sun.SciRep.2016}: 
\begin{equation}
	\mathrm{CI} (S^*,Q^*) = \frac{\Gamma_{S^*}(S^*) + \Gamma_Q^*(Q^*) - \Gamma_{S^*}(S^*) \cdot \Gamma_Q^*(Q^*)}{\Gamma_{S^*Q^*}(S^*,Q^*)},
\end{equation}
where $\Gamma_{S^*}(S^*) = \frac{R_i(0,0) - R_i(S^*,0)}{R_i(0,0)}$ represents the relative change in tumor reduction due to the antigen alone. $\Gamma_{Q^*}(Q^*) = \frac{R_i(0,0) - R_i(0,Q^*)}{R_i(0,0)}$ indicates the relative change in tumor reduction due to adjuvant alone. $\Gamma_{S^*}(S^*) + \Gamma_Q^*(Q^*) - \Gamma_{S^*}(S^*) \cdot \Gamma_Q^*(Q^*)$ denotes the expected effect of the treatment. $\Gamma_{S^*Q^*}(S^*,Q^*) = \frac{R_i(0,0) - R_i(S^*,Q^*)}{R_i(0,0)}$ indicates the simulated effect of the treatment. $\mathrm{CI} (S^*,Q^*)<1$ represents a synergistic effect between the antigen and the adjuvant, or else an antagonistic effect. The calculation showed that the antigen and adjuvant exhibit synergistic effects in the dose space (Fig. \ref{Fig10}F). At low adjuvant doses, the antigen and adjuvant exhibit a high degree of synergy (Fig. \ref{Fig10}F). Conversely, at low antigen levels, antigen and adjuvant exhibit lower synergy (Fig. \ref{Fig10}F). 

\begin{figure}[h!]
	\centering
	\includegraphics[width=16cm]{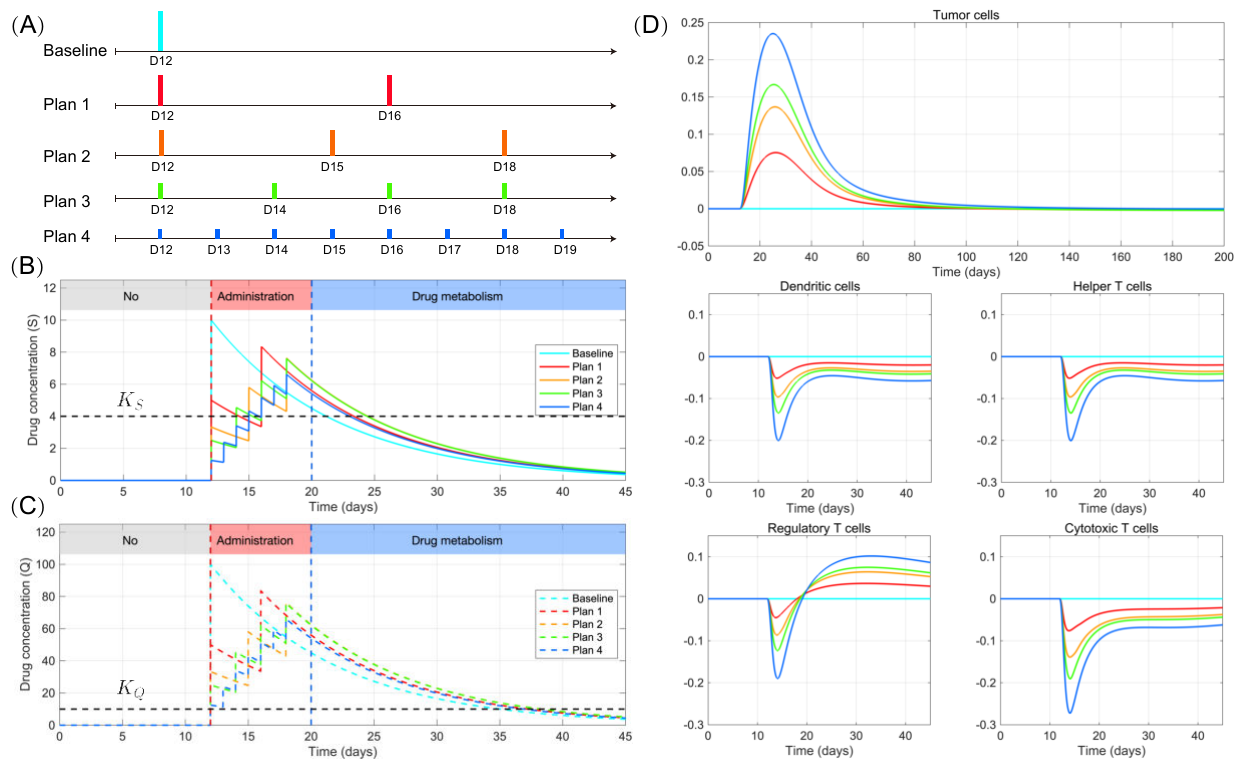}
	\caption{\textbf{Pharmacodynamic evaluation of cancer vaccines under different treatment plans.} (A) Administration time and dose design for different treatment plans. Plan 1: 4-day interval therapy; Plan 2: 3-day interval therapy; Plan 3: 2-day interval therapy; Plan 4: 1-day interval therapy. All four plans maintain the same total dosage as the baseline plan over a 12-19 day treatment period. (B)-(C) PK curves of antigen and adjuvant in cancer vaccines. (D) Relative changes in $C$, $D$, $T_h$, $T_r$, and $T_c$ under different plans relative to the baseline plan (cyan curve). Parameter values are given in Table \ref{Tab:Parameter}.}
	\label{Fig11}
\end{figure}

To evaluate the effect of different cancer vaccine treatment plans on the evolution of tumor-immune system dynamics, we developed 4 differentiated dosing plans and performed a comparative study (Fig. \ref{Fig11}A). The results showed that multiple low-dose plans (Plans 1-4) can increase antigen and adjuvant exposure levels during metabolism compared to baseline treatment (Fig. \ref{Fig11}B and C). However, a low initial dose will result in insufficient activation of $D$, which will weaken the activation of $T_h$ and $T_c$ (Fig. \ref{Fig11}D). Meanwhile, this also leads to amplification of $C$ and $T_r$ (Fig. \ref{Fig11}D). These results showed that the efficacy of the cancer vaccine was dose-dependent, and the strength of the anti-tumor immune response was positively correlated with the injection dose. This means that the maximum dose therapy of the cancer vaccine will have a stronger anti-tumor effect in clinical practice.

\subsection{Tumor-immune heterogeneity modeling and prognostic analysis}

In tumor heterogeneity modeling (Section 4.1), we considered tumor and pharmacodynamic variability, but fixed immune-related parameters. This meant that we neglected immune heterogeneity. To further integrate immune heterogeneity into the modeling framework, we selected six key parameters characterizing inter-individual variations in baseline immune cell levels and target protein expression: (1) immature dendritic cells ($D_0 \in [1\times10^7,5\times10^7]$); (2) na\"{i}ve CD4+ T cells ($T_{N4} \in [1\times10^9,5\times 10^9]$); (3) na\"{i}ve CD8+ T cells ($T_{N8} \in [1\times10^9,5\times 10^9]$); (4) dendritic cell half-saturation constant ($K_D\in [3\times10^8,8\times10^8]$); (5) PD-L1 expression rate ($\rho_L \in [1.2\times10^{-6},3.2\times10^{-6}]$); (6) PD-1 expression rate ($\rho_P \in [6\times10^{-7},1.2\times10^{-6}]$). We constructed an ensemble representing tumor-immune heterogeneity by combining 300 randomly sampled parameter sets from six immunological parameter ranges with the top 300 ABC-selected sets based on goodness-of-fit. This ensemble was used to simulate four treatment strategies across 300 virtual cohorts.

\begin{figure}[h!]
	\centering
	\includegraphics[width=16cm]{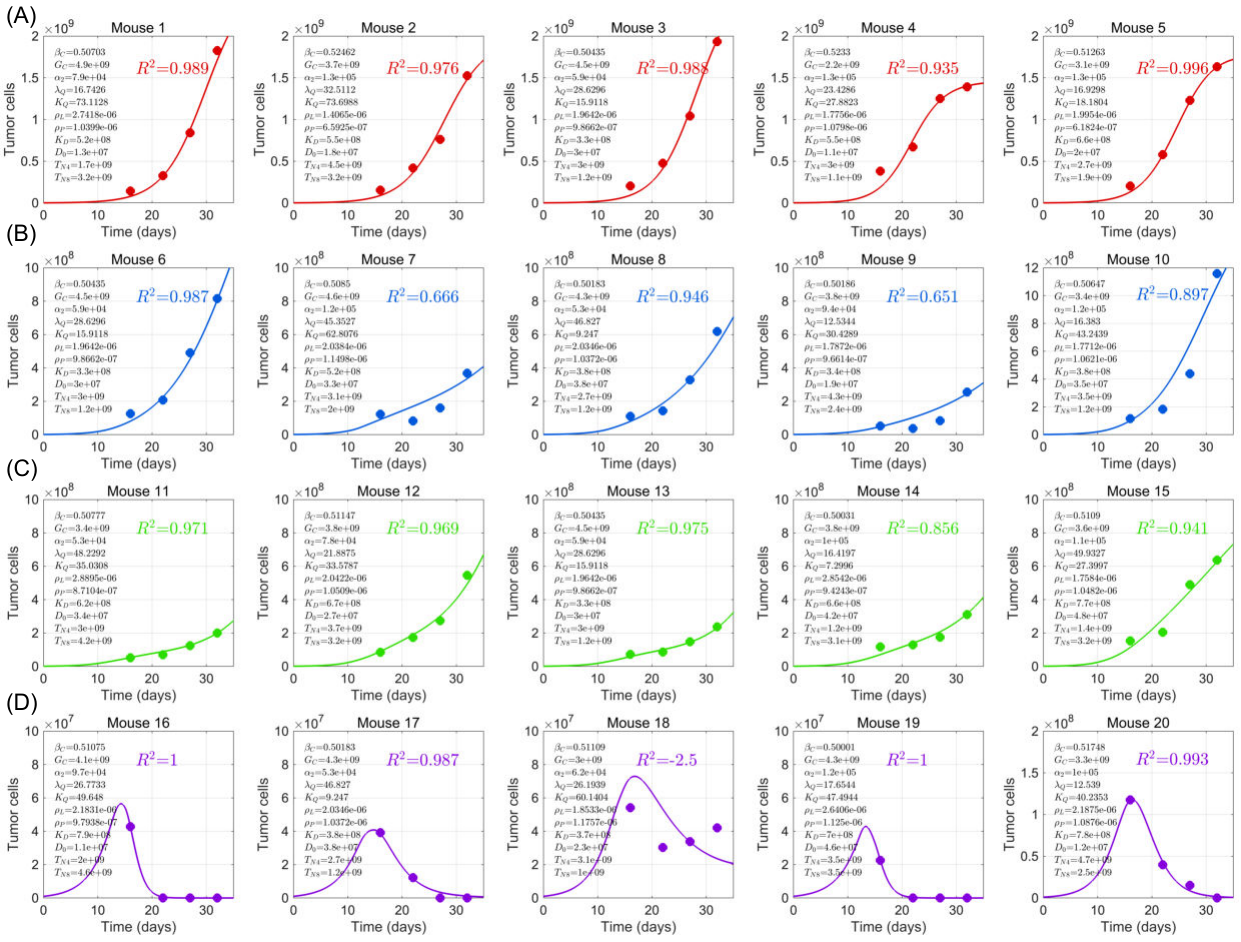}
	\caption{\textbf{Individual best-fit dynamics of the tumor-immune heterogeneity model by 300 virtual cohorts.} (A) $\sim$ (D) Individualized treatment response analysis: Tumor growth dynamics of 20 mice in control (A), cancer vaccine (B), anti-PD-L1, (C) and combination therapy (D) groups. The lines represent the best-fit model trajectories for 300 virtual cohorts. The text in the upper left corner indicates the corresponding parameter values in the tumor-immune heterogeneity modeling framework.}
	\label{Fig12}
\end{figure}

Computational results demonstrated strong concordance between experimental tumor growth curves and model simulations across treatment groups (Fig. \ref{Fig12}). Specifically:
\begin{itemize}	
	\item \textbf{Control group} exhibited $R^2$ values of 0.989, 0.976, 0.988, 0.935 and 0.996 (Fig. \ref{Fig12}A).
	\item \textbf{Cancer vaccine group} showed $R^2$ values of 0.987, 0.666, 0.946, 0.651 and 0.897 (Fig. \ref{Fig12}B).
	\item \textbf{Anti-PD-L1 group} demonstrated $R^2$ values of 0.971, 0.969, 0.975, 0.856 and 0.941 (Fig. \ref{Fig12}C).
	\item \textbf{Combination therapy group} displayed $R^2$ values of 1.000, 0.987, -2.500, 1.000 and 0.993 (Fig. \ref{Fig12}D).
\end{itemize}
Notably, mouse 9 (vaccine group) achieved $R^2$ improvement from -0.995 to 0.561 (Fig. \ref{Fig4}D vs. \ref{Fig12}B). Meanwhile, mouse 18 (combination therapy) achieved $R^2$ improvement from -4.760 to -2.500 (Fig. \ref{Fig4}F vs. \ref{Fig12}D). This suggests that the tumor-immune heterogeneity model can more accurately capture the dynamic features of abnormal samples, validating the advantages of the model in characterising the heterogeneity of complex immune microenvironments. Additionally, we observed that the vaccine treatment data exhibit an initial regression phase, which was not captured in the simulations of any virtual patient cohort.

\begin{figure}[h!]
	\centering
	\includegraphics[width=16cm]{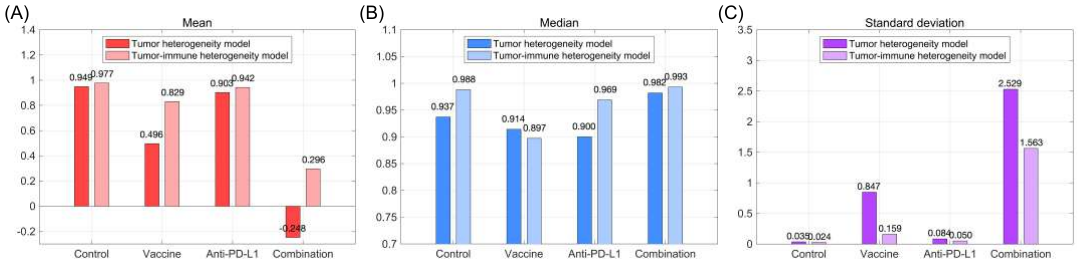}
	\caption{\textbf{Statistical analysis of $R^2$ in the tumor heterogeneity modeling framework and the tumor-immunity heterogeneity modeling framework.} (A) Mean. (B) Median. (C) Standard deviation. }
	\label{Fig13}
\end{figure}

To further compare the difference in data fitting performance between the tumor heterogeneity and the tumor-immunity heterogeneity modeling framework, we compared the statistical characteristics of the coefficient of determination ($R^2$): mean, median, and standard deviation. The results showed that the mean value of $R^2$ was significantly increased under the tumor-immunity heterogeneity modeling framework, 0.949 vs. 0.977 (control), 0.496 vs. 0.829 (vaccine), 0.903 vs. 0.942 (anti-PD-L1), and -0.248 vs. 0.296 (combination) (Fig. \ref{Fig13}A). Meanwhile, the median was also elevated in all groups except the cancer vaccine group (0.914 vs. 0.897), 0.937 vs. 0.988 (control), 0.900 vs. 0.969 (anti-PD-L1), and 0.982 vs. 0.993 (combination) (Fig. \ref{Fig13}B). The standard deviation was significantly reduced, 0.035 vs 0.024 (control), 0.847 vs 0.159 (vaccine), 0.084 vs 0.050 (anti-PD-L1), and 2.529 vs 1.563 ( combination) (Fig. \ref{Fig13}C). The above results indicate that integrating immune heterogeneity features not only improves model accuracy (increasing mean and median), but also enhances predictive stability (decreasing standard deviation). This finding provides an important basis for the subsequent study of the quantitative relationship between immune cell dynamics and tumor dynamics.

\begin{figure}[h!]
	\centering
	\includegraphics[width=16cm]{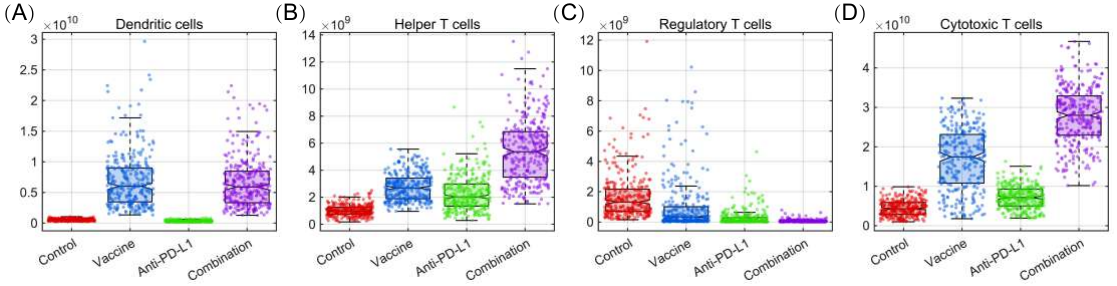}
	\caption{\textbf{Distribution of immune cells under different treatment strategies.} Distribution of $D$, $T_h$, $T_r$, and $T_c$ in the control, cancer vaccine, anti-PD-L1 treatment, and combination therapy groups on day 45.}
	\label{Fig14}
\end{figure}

To elucidate the immune dynamics under different treatment strategies, we conducted numerical simulations across 300 virtual cohorts incorporating tumor-immune heterogeneity, with particular emphasis on analyzing the distribution of four immune cell populations at day 45. The results showed that the cancer vaccine and combination therapy significantly promoted dendritic cells expansion, while anti-PD-L1 monotherapy could not increase the number of dendritic cells (Fig. \ref{Fig14}A). This could be related to the fact that cancer vaccines promote dendritic cells' maturation. In terms of adaptive immunity, all treatment groups increased the level of helper T cells, with the combination therapy having the most significant effect (Fig. \ref{Fig14}B). Cancer vaccines enhance cytotoxic T cells more than anti-PD-L1 treatment (Fig. \ref{Fig14}C). In addition, cancer vaccines are able to reduce regulatory T cell levels, and anti-PD-L1 and combination therapy are able to further reduce regulatory T cell levels (Fig. \ref{Fig14}D). Furthermore, the experimental results further demonstrate that recruitment of helper T cells and cytotoxic T cells contributes to control tumor growth and improved survival rates \cite{Erdag.CancerRes.2012,Davidsson.ModPathol.2013,Weigelin.NatCommun.2021}. Conversely, tumor infiltration by regulatory T cells is generally indicative of immune escape \cite{Facciabene.CancerRes.2012}. These results demonstrate that the combined cancer vaccine and anti-PD-L1 treatment synergistically enhances immune activation and overcomes immunosuppressive mechanisms, effectively driving the essential immune cascade from antigen presentation through effector cell activation to immunosuppression reversal.

To evaluate the predictive value of immune cell subsets on tumor progression, we analyzed the predictive efficacy of five key immune indexes ($T_h$, $T_r$, $T_c$, $T_h/T_r$, and $T_c/T_r$) using a binary classification model in machine learning. The model used tumor cell count ($C$) at day 45 as the outcome variable, classifying patients into responders and non-responders based on the sample mean. Predictors included immune cell counts ($T_h$, $T_r$, $T_c$) at days 7 and 21 to capture immune dynamics before and after treatment. By correlating these immune markers with tumor burden, key immunological features predictive of tumor progression were identified, providing a quantitative basis for dynamic monitoring and prognostic assessment in immunotherapy. The results showed that $T_c$ was the most significant predictive biomarker before anti-PD-L1 treatment (AUC = 0.97), followed by $T_c/T_r$ (AUC = 0.88). After anti-PD-L1 treatment, the AUC of $T_c$ was further increased to 0.99, followed by $T_h/T_r$ (AUC = 0.86) and $T_c/T_r$ (AUC = 0.85). Before the cancer vaccine treatment, $T_c/T_r$ was the most significant predictive biomarker (AUC = 0.95). This indicates that the combined index of $T_c$ and $T_r$ can effectively predict tumor progression. After cancer vaccine treatment, $T_c$ was the most significant predictive biomarker (AUC = 0.99). Before combination therapy, $T_c/T_r$ was the most significant predictive biomarker (AUC = 0.88). After combination therapy, $T_c$ was the most significant predictive biomarker (AUC = 0.99). These results demonstrated that $T_c$ is a stable predictive biomarker in various therapeutic strategies. $T_c/T_r$ is an important predictive biomarker before therapy. 

\begin{figure}[h!]
	\centering
	\includegraphics[width=16cm]{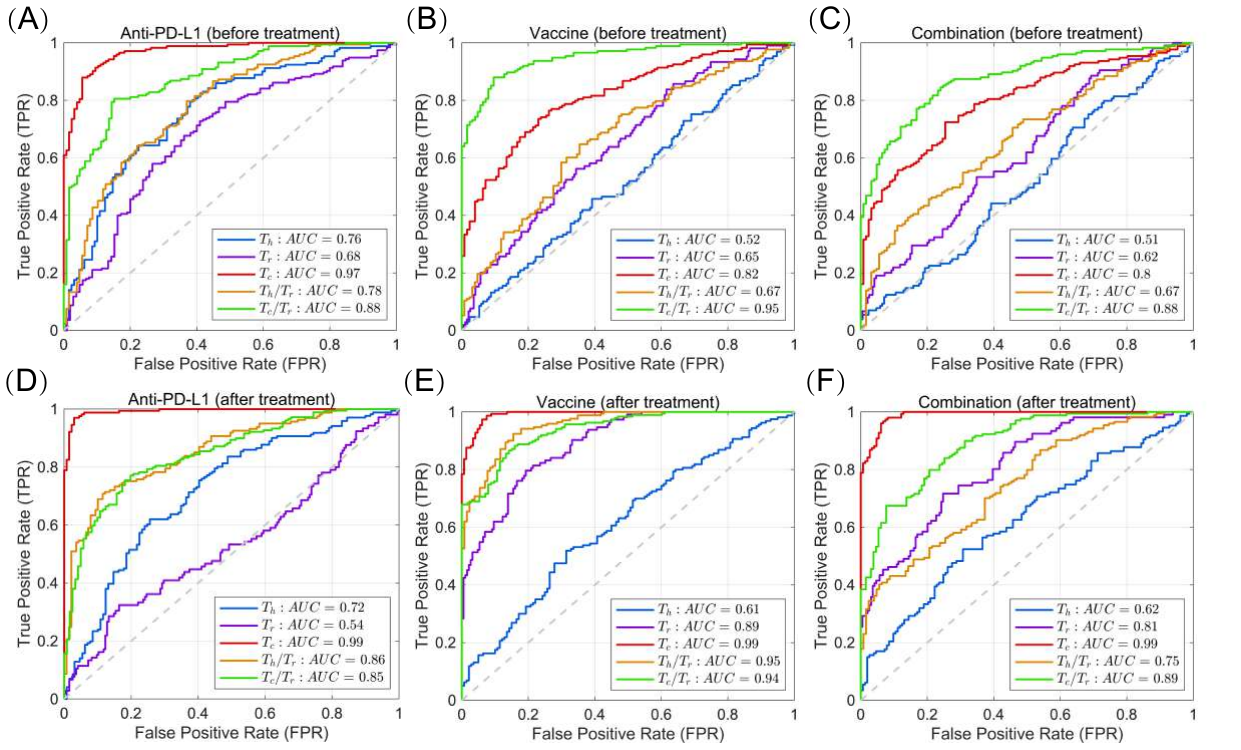}
	\caption{\textbf{ROC analysis of predictive biomarkers.} (A) $\sim$ (C) ROC analysis of predictive biomarkers before anti-PD-L1 therapy, cancer vaccines and combination therapy. (D) $\sim$ (F) ROC analysis of predictive biomarkers after anti-PD-L1 therapy, cancer vaccines and combination therapy. ROC: receiver operating characteristic. AUC: area under the curve. }
	\label{Fig15}
\end{figure}

\section{Discussion}

Mathematical modeling has emerged as a pivotal tool for characterizing tumor evolutionary dynamics \cite{Eftimie.BullMathBiol.2016,Eftimie.BullMathBiol.2023,Li.CSIAM-LS.2025}. Particularly, its capacity to quantitatively map tumor-immune regulatory networks provides a systems biology perspective for understanding oncogenesis. In this study, we developed a multi-scale mathematical model integrating cell population dynamics, cytokine regulatory networks, and pharmacological characteristics to elucidate the complex dynamics of tumor-immune interactions under anti-PD-L1 therapy and cancer vaccines. The results showed that: (1) multiple low-dose administrations of anti-PD-L1 significantly reduce tumor burden compared to baseline treatment; (2) the maximum-dose administration of the cancer vaccine showed optimal dynamics in tumor growth inhibition; and (3) cytotoxic T cells demonstrated predictive utility both pre- and post-treatment, whereas the cytotoxic T cells-to-regulatory T cells ratio specifically emerged as a pre-treatment predictive biomarker. Meanwhile, to analyze the different evolution of tumor data under the same treatment strategy, we created a virtual sample cohort based on an ABC method to capture the treatment differences between individuals. The results suggest that a modeling framework that integrates immune heterogeneity is more beneficial to accurately capture the dynamics of tumor evolution.

In previous research, we focused on mechanistic modeling of combined anti-FGFR and anti-PD-1 therapy, emphasizing the interactions between molecular pathways under targeted and immune-combination interventions \cite{Li.BullMathBiol.2024}. While both the previous study and the current one aim to understand the synergistic effects of combination therapy, this study uses ABC method to create personalized digital twins, which significantly improves the characterization of heterogeneity between patients. Furthermore, we previously developed a quantitative cancer-immunity cycle model to predict disease progression in advanced colorectal cancer \cite{Li.NPJSystBiolAppl.2025}. While both studies adopt a systems modeling approach and utilise virtual patient cohorts to capture population heterogeneity, our earlier work relied on conventional sampling methods for parameter generation and did not incorporate probability inversion algorithms, such as ABC. Consequently, it offered limited flexibility when it came to handling high-dimensional uncertainty and individualised parameter inference. In summary, the current framework provides a more interpretable, flexible, and clinically predictive modeling platform. This establishes a more reliable computational foundation for optimizing immunotherapy and informing personalised treatment strategies.

Mechanistic modeling and data-driven approaches offer complementary strengths in cancer treatment. While data-driven methods leverage large datasets to predict drug efficacy, they often lack interpretability and generalize poorly beyond available data. In contrast, mechanistic modeling provides high interpretability and enables dynamic simulation of biological processes. In the future, dual-driven research integrating both data and mechanistic models is expected to become a major focus. Mechanistic models can act as digital twins to generate synthetic data for training reinforcement learning agents, while data-driven techniques can refine model parameters and optimize structures. This combined framework is poised to enhance the robustness and interpretability of AI models, accelerating advances in precision medicine. Recently, Liu et al. \cite{Liu.SciAdv.2025} introduced a multiscale model-driven reinforcement learning framework to simulate tumor-immune interactions and optimize combination therapies. Their work provides a groundbreaking solution to key challenges in precision oncology.

This study employs a multiscale ODE framework to simulate tumor-immune interactions and drug response dynamics. The model assumes a uniformly mixed cell population, which does not fully capture the spatial heterogeneity and complex structural features of the actual tumor microenvironment. It is essential to recognize that factors such as spatial cell distribution, local concentration gradients, physical barriers, and heterogeneous cell contacts have a significant impact on key biological processes, including immune cell infiltration efficiency, targeted therapy response, and the development of drug resistance. Due to the neglect of spatial effects, the current model may overestimate the diffusion rate and therapeutic efficacy of immune modulators and oversimplify the dynamics of cellular interactions. In future studies, incorporating spatially explicit modeling approaches such as ABM or PDE frameworks could enhance the model's predictive power and biological realism. Recently, Lin et al. \cite{Lin.MultiscaleModelSim.2025} developed a coupled ODE-PDE multiscale modeling framework and a hybrid analytical-numerical approach, revealing the mechanism of resistance to CSF1R inhibitors in glioblastoma. Their work establishes a critical foundation for deciphering spatial heterogeneity in the tumor immune microenvironment.

The present modeling framework is designed to simulate system-level dynamics in combination with cancer therapy, yet it omits more detailed biological details. Immune cell populations are represented as functionally homogeneous compartments, thereby failing to fully capture processes such as T cell exhaustion, memory subset differentiation, or drug resistance mechanisms. This simplified approach prioritizes a balance between model complexity and interpretability, ensuring mathematical and computational tractability while mitigating the risk of over-parameterization and loss of robustness. By maintaining this coarse-grained resolution, the model focuses on core system-level interactions among tumor cells, key immune populations, and treatment interventions. Nevertheless, this simplification constrains the ability to quantitatively characterize specific mechanistic behaviors or predict finely resolved phenotypic outcomes. Rapidly evolving technologies such as single-cell sequencing, spatial transcriptomics, and multi-omics are now providing rich data to help bridge this gap. Future work will leverage these resources to extend the model to incorporate cellular states and spatial resolution. Integrating these details is essential for developing predictive digital twin models of tumor-immune interactions.

A further limitation of the current model is its neglect of hypoxic regions within the tumor microenvironment. It is well established that hypoxia exerts profound immunosuppressive effects, such as impairing T cell function, promoting the recruitment of immunosuppressive cells, and upregulating immune checkpoint molecules, including PD-L1. The absence of hypoxia-driven mechanisms may lead to an overestimation of immune activation and anti-tumor efficacy, particularly in models simulating immune checkpoint inhibitors or vaccine-based therapies. Future iterations of the model would benefit from incorporating oxygen gradients and their molecular and cellular consequences, thereby enabling more accurate predictions of spatial immune suppression and therapy resistance. Such an extension would provide deeper mechanistic insights into the combined effects of hypoxia and immunotherapy within a multiscale framework.

In this study, we employed ABC method to generate a diverse virtual patient cohort that effectively captures inter-individual heterogeneity in response to anti-PD-L1 and cancer vaccine combination therapy. By integrating prior knowledge with empirical data, the ABC-based calibration allowed us to infer patient-specific parameters and simulate realistic inter-patient variability in immune-tumor dynamics. This approach not only enhances the biological plausibility of the model but also provides a robust computational framework for generating in silico digital twins that mirror the clinical diversity of real-world patient populations. These digital twins enable the exploration of individualized treatment responses and the identification of subpopulations that may benefit most from specific therapeutic interventions. Additionally, the integration of optimization algorithms into this ABC-powered digital twin platform offers a promising pathway toward truly personalized treatment strategies. For example, reinforcement learning or Bayesian optimization techniques could be applied to dynamically adjust drug dosing schedules and combination regimens based on continuously updated, patient-specific in silico feedback. By iteratively refining model parameters using longitudinal clinical data and incorporating multi-omics inputs, the digital twin framework can mature into a robust and clinically relevant platform for precision immuno-oncology. This approach not only enhances individual-level predictive capability but also provides a scalable methodology for optimizing adaptive combination immunotherapies in the context of high-dimensional biological heterogeneity.

\section*{CRediT authorship contribution statement}

\textbf{Chenghang Li:} Conceptualization, Investigation, Methodology, Software, Writing – review \& editing, Writing – original draft, Visualization. \textbf{Haifeng Zhang:} Methodology, Investigation, Writing – review \& editing, Writing – original draft. \textbf{Xiulan Lai:} Conceptualization, Writing – review \& editing, Writing – original draft, Supervision, Project administration, Funding acquisition. \textbf{Jinzhi Lei:} Conceptualization, Methodology, Resources, Writing – review \& editing, Writing – original draft, Supervision, Project administration, Funding acquisition. 

\section*{Declaration of competing interest}

The authors declare that they have no conflicts of interest.

\section*{Data availability}

The data used in this article can be accessed in published studies \cite{Liu.NatCancer.2022}.

\section*{Acknowledgments}

This work was supported by the Key Programme of the National Natural Science Foundation of China (No. 12331018 to J. Lei and X. Lai), General Program of the National Natural Science Foundation of China (No. 12171478 to X. Lai).

\section*{Appendix. Parameter estimation}\label{Appendix1}

The model parameter estimation followed a structured, multi-stage approach to ensure robustness and mitigate the challenges of high-dimensional parameter identification. Model calibration was primarily performed using the average tumor volume dynamics from four treatment groups (control, vaccine, anti-PD-L1, and combination therapy) in C57BL/6J mice bearing MC38 tumors. The coefficient of determination was used as the loss function to evaluate the relationship between simulated results and experimental measurements. Parameter optimization was carried out using the Markov Chain Monte Carlo (MCMC) algorithm. The specific steps were as follows.

\textbf{Step 1: Literature-based parameter fixation.} A total of 40 parameters are directly assigned values derived from established biological experiments or previously published models. These parameters were kept fixed throughout subsequent calibration steps. These values and their literature sources were described as follows:
\begin{itemize}
	\item (1) Initial values for the tumor-immune system. Based on previous studies by Li et al. \cite{Li.BullMathBiol.2024} and Chen et al. \cite{Chen.MathBiosci.2022}, the order of magnitude of the initial value of immune cells was determined to be $ 10^8$ cells. The initial value of tumor cells was determined to be $1 \times 10^6$ cells based on mouse experiments \cite{Liu.NatCancer.2022}.
	\item  (2) Number of immature or na\"{i}ve cells. Based on previous studies \cite{Chen.MathBiosci.2022,Rodriguez-Messan.2021.PLoSComputBiol}, we referred the number of immature dendritic cells as $1.94 \times 10^7$ cells, and the number of na\"{i}ve CD4+ and CD8+ T cells as $3.77 \times 10^9$ cells and $1.61 \times 10^9$ cells, respectively.
	\item (3) Proliferation rate ($\beta_x$). According to the studies \cite{Lai.SciChinaMath.2020,Friedman.BullMathBiol.2018,Pillis.CancerRes.2005,Sardar.CommunNonlinearSci.2023}, the proliferation rates of immune cells and tumor cells were selected as 0.25 days$^{-1}$ and 0.514 days$^{-1}$, respectively. 
	\item (4) Activation rate ($\lambda_x$). We referred to studies \cite{Li.BullMathBiol.2024,Rodriguez-Messan.2021.PLoSComputBiol,Lai.SciChinaMath.2020,Wang.JImmunotherCancer.2021} that determined the order of magnitude of cell activation rates at $10^0$ - $10^2$ day$^{-1}$.
	\item (5) Death rate ($d_x$). We referred to Li et al. \cite{Li.BullMathBiol.2024} and Anbari et al. \cite{Anbari.NPJSystBiolAppl.2024}, where the magnitude of cell death rate was determined to $10^{-2}$ - $10^{-1}$ day$^{-1}$. For immune cells, $d_x$ represents not only death but also functional loss.
	\item (6) Half-saturation constant ($K_x$) and suppression function ($K_{xy}$). We referred to \cite{Li.BullMathBiol.2024,Robertson-Tessi.JTheorBiol.2012,Zhang.TheoryBiosci.2025} to select the half-saturation constants and inhibitory functions of the cytokines in the magnitude of $10^{-1}$ - $10^{2}$ ng mL$^{-1}$.
	\item (7) Production rate ($\delta_{xy}$). Based on the research \cite{Li.BullMathBiol.2024,Robertson-Tessi.JTheorBiol.2012,Zhang.TheoryBiosci.2025}, we selected the magnitude of cytokine production rate as $10^{-10}$ - $10^{-7}$ ng mL$^{-1}$ day$^{-1}$ cell$^{-1}$. Since the secretion rate of $I_2$ by $T_c$ is much lower than that by $T_h$, we set $\delta_{I_2T_c}=\frac{1}{50}\delta_{I_2T_h}$.
	\item (8) Degradation rate ($d_y$). We referred to \cite{Lai.PNAS.2018,Li.BullMathBiol.2024,Rodriguez-Messan.2021.PLoSComputBiol,Lai.SciChinaMath.2020,Friedman.BullMathBiol.2018,Qomlaqi.MathBiosci.2017,Liao.MathBiosci.2023}, the magnitude of the degradation rate of cytokines and tumor antigens was determined at $10^{0}$ - $10^{2}$ day$^{-1}$.
	\item (9) Tumor growth and pharmacodynamic (PD) parameters. The carrying capacity ($G_C$) of the tumor was selected as $3.00 \times 10^{9}$ cells \cite{Qomlaqi.MathBiosci.2017,Ndenda.ChaosSolitonFract.2021}. The Hill coefficient ($n$) was selected as $3$. The expression rates of PD-1 and PD-L1 ($\rho_P$ and $\rho_L$) were chosen to be $1 \times 10^{-6}$ and $2.5 \times 10^{-6}$ nmol L$^{-1}$ cell$^{-1}$ \cite{Li.BullMathBiol.2024}, respectively. The amplification factor of PD-L1 was $50$ \cite{Li.BullMathBiol.2024}. The equilibrium constant for PD-1-PD-L1 was chosen to be $50$ L nmol$^{-1}$. 
\end{itemize}

\textbf{Step 2: Estimation of cytokine half-saturation constant and suppression function.} We established the value ranges for the half-saturation constant and inhibition function based on existing literature (see Step 1).  The half-saturation constants and inhibition functions were iteratively adjusted based on the tumor growth data of the control group, with consideration given to the model output, parameter ranges, and biological relevance. During this process, we generated time-series tumor evolution data by interpolating the tumor measurements and incorporated them into the model simulations, thereby eliminating the need for additional estimation of tumor-related parameters.

\textbf{Step 3: Identification of tumor-related parameters.} Using the control group data and the parameter values from Steps 1 and 2 as priors, we estimated tumor-specific parameters including $\delta_{SC}$, $K_D$, $n$, $\eta_{T_c}$, $\eta_{T_h}$, and $K_{P_L}$. Biologically plausible ranges were enforced for all parameters during MCMC sampling.

\textbf{Step 4: Calibration of vaccine-related pharmacodynamic parameters.} Using data from the vaccine treatment group, we estimated the following parameters specific to the vaccine response: $\lambda_Q$, $K_S$, $K_Q$, $\mu_S$, and $\mu_Q$. To ensure parameter physiological plausibility, we employed initial sampling ranges based on the biological meanings of these parameters, and the MCMC method was used for further identification.

\textbf{Step 5: Calibration of anti-PD-L1 pharmacodynamic parameters.} Data from the anti-PD-L1 monotherapy group were used to estimate parameters related to immune checkpoint inhibition ($\alpha_2$, $\mu_A$). Their initial sampling ranges were determined based on known pharmacodynamic principles and earlier model behavior.

Finally, all parameters were tested using a cross-validation strategy that incorporated the combination therapy group, which had not been used in prior steps. This approach ensured the model's generalizability to unseen conditions and confirmed its predictive utility, particularly in recapitulating complex treatment interactions.

\bibliographystyle{unsrt} 
\bibliography{references}

\end{document}